\documentclass[12pt]{iopart}
\usepackage{iopams}
\usepackage{bbm}
\usepackage{graphicx}
\usepackage{esvect}
\usepackage{color,cite}
\usepackage{bm}

\newenvironment{remark}[1][Remark:]{\begin{trivlist}
\item[\hskip \labelsep {\bfseries #1}]}{\end{trivlist}}

\begin{document}

\title[Current fluctuations in a semi-infinite line]{Current fluctuations in a semi-infinite line}

\author{Soumyabrata Saha and Tridib Sadhu$^\star$}

\address{Department of Theoretical Physics, Tata Institute of Fundamental Research, Homi Bhabha Road, Mumbai 400005, India}

\ead{$^\star$tridib@theory.tifr.res.in}

\begin{abstract}
    We present the application of a fluctuating hydrodynamic theory to study current fluctuations in diffusive systems on a semi-infinite line in contact with a reservoir with slow coupling. We show that the distribution of the time-integrated current across the boundary at large times follows a large deviation principle with a rate function that depends on the coupling strength with the reservoir. The system exhibits a long-term memory of its initial state, which was earlier reported on an infinite line and can be described using quenched and annealed averages of the initial state. We present an explicit expression of the rate function for independent particles, which we verify using an exact solution of the microscopic dynamics. For the symmetric simple exclusion process, we present expressions for the first three cumulants of both quenched and annealed averages.
\end{abstract}


\submitto{J. Stat. Mech.}

\maketitle

\section{Introduction}
Non-equilibrium processes are ubiquitous in nature. However, unlike equilibrium statistical mechanics, no universal framework exists to study macroscopic fluctuations for out-of-equilibrium systems. In the past two decades a powerful hydrodynamic theory, commonly referred as the macroscopic fluctuation theory (MFT), has emerged \cite{bertini_macroscopic_2015,derrida_microscopic_2011} that presents a systematic approach for studying fluctuations in diffusive transport models. Since its inception in early 2000 by Bertini, De Sole, Gabrielli, Jona-Lassinio, and Landim \cite{bertini_fluctuations_2001,bertini_macroscopic_2002}
the theory has been successfully applied in a varied range of non-equilibrium scenarios. Notably, the exact results for the large deviations of density \cite{derrida_large_2002} and current \cite{bodineau_current_2004,Derrida2004Roche} in the non-equilibrium stationary state of the symmetric exclusion process on a finite line have been independently reproduced \cite{bertini_macroscopic_2002,tailleur_mapping_2007,bertini_current_2005,Lecomte_Current_2010} within this hydrodynamic approach. On an infinite line, the highly non-trivial Bethe ansatz result \cite{Gerschenfeld2009Bethe} about current fluctuations in the symmetric simple exclusion process has been recently reproduced \cite{mallick2022exact} using the MFT formalism \cite{derrida_current_2009} by discovering a remarkable connection with an integrable model. Similar exact results about large deviations in related transport models \cite{bertini_large_2005}, tracer statistics in single-file \cite{Krapivsky2014,Krapivsky2015}, height fluctuations in KPZ \cite{Krajenbrink2021,Meerson2016} have made the MFT a reliable theoretical approach.

For the exclusion process the two commonly studied geometry are finite one dimensional lattice coupled with two unequal reservoirs at the boundary and infinite lattice starting with a non-stationary state. The two geometry represents two different non-equilibrium scenarios. A finite system reaches a non-equilibrium stationary state driven by the unequal particle reservoirs at the two ends where the initial state of the system becomes irrelevant. On the other hand, an infinite system started with a step initial density profile takes infinite time to reach the asymptotic equilibrium state. For the latter, it was found \cite{derrida_current_2009} that the initial state plays a determining factor for the fluctuations even at large times.

In this paper, our interest is in the intermediate geometry: a semi-infinite system coupled with a reservoir at one end. When the initial density of the system is different from the reservoir density, the system is out of equilibrium and evolves to reach the asymptotic stationary state in equilibrium with the reservoir. Studies of this geometry are significantly scarce. Earliest work about exclusion process on semi-infinite lattice is by Ligget \cite{Liggett_1975} and then by Grosskinsky \cite{Grosskinsky_thesis}. For the asymmetric exclusion process, Sassamoto and Williams showed \cite{Williams2000} that there is a convergence to a local stationary state where spatial correlation of density has an exact correspondence with the correlation in the finite geometry. Recently Duhart et al \cite{Duhart_2018} derived the large deviations function for the empirical density in the local stationary state. There is an exact result \cite{Tracy_2013} about transition probability that has been exploited to obtain the mean and the variance of integrated current for the symmetric exclusion process for an initial state with Bernoulli measure. These exact results about the two cumulants are related to the cumululants of current on an infinite line coupled to a reservoir at the centre \cite{Krapivsky2012}. 

A reason for us working with the semi-infinite geometry is that the effect of reservoir on the boundary layer can be studied for the non-stationary state. At the same time we are able to study how the system keeps a long-term memory of its initial state in presence of the reservoir.
\begin{figure}
\centering \includegraphics[width=0.90\textwidth]{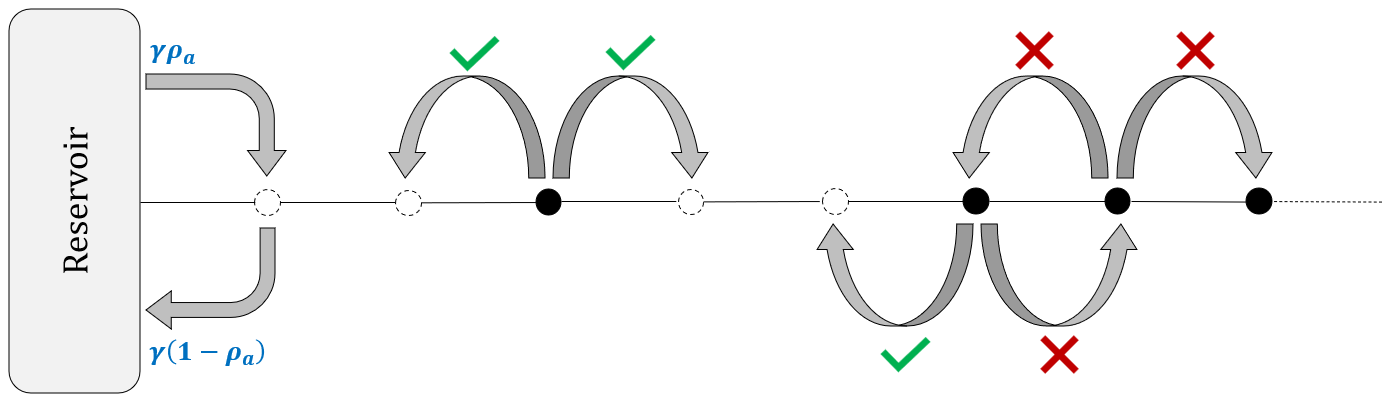}
    \caption{The symmetric simple exclusion process on a one dimensional semi-infinite lattice coupled with the boundary reservoir of density $\rho_a$, where $\gamma$ represents the strength of the jump rates. In the bulk, jump rates are of unit strength.}
    \label{fig: SSEP Slow}
\end{figure}

In order to understand the effect of the reservoir we introduce a tuning parameter $\gamma>0$ for the jump rates to and from the reservoir. A large $\gamma\gg 1$ implies a fast time scale at the boundary compared to the bulk dynamics and one would expect the latter to be the determining factor for transport. For small $\gamma\ll 1$, the slow boundary rates are the bottleneck. The question we ask is how different are the fluctuations in these two regimes and whether there is a scale of $\gamma$ that demarcates the two regimes.

For a quantitative study of these questions, we consider two simple examples: particles hopping on a semi-infinite $\mathbb{Z}^+$ lattice (a) without an inter-particle interaction (b) and with onsite exclusion interactions. The second example is the symmetric simple exclusion process (SSEP) whose dynamics is sketched in \fref{fig: SSEP Slow}. The jump rates at the boundary correspond to a reservoir of density $\rho_a$ independent of the parameter $\gamma$ which only controls the time-scale compared to the bulk dynamics. Initially the system is at uniform average density $\rho_b$. The low-density limit of the SSEP corresponds to the first example of non-interacting particles. 

\begin{figure}
\centering
    \includegraphics[width=0.75\textwidth]{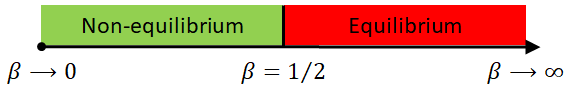}
    \caption{For the SSEP in \fref{fig: SSEP Slow} with $\gamma\sim T^{-\beta}$, the marginal value $\beta=1/2$ separates the slow coupling  and the fast-coupling regimes of the current fluctuations. In the former regime the system is effectively in equilibrium. }
    \label{fig: gamma}
\end{figure}

We specifically look at the large time statistics of the time-integrated current $Q_T$ across the system-reservoir boundary which gives the net influx of particles in total time $T$ from the reservoir to the system. For a diffusive system like the two examples in consideration, the relevant scale for $Q_T$ is $\sim \sqrt{T}$. We show that $\gamma \sim T^{-1/2}$ is the marginal scale of the boundary rates that separates the slow and the fast coupling regimes of current fluctuation. More accurately, for $\gamma\sim T^{-\beta}$ with $\beta<1/2$ the long time fluctuations of $Q_T$ are independent of $\gamma$ and are same as for the infinitely fast reservoir coupling. Similar behaviour was noted in \cite{Krapivsky2012} for SSEP on a different geometry. In the slow coupling regime of $\beta>1/2$ the fluctuations are governed by the single bond between the reservoir and the system which is effectively in equilibrium. This behaviour is illustrated in \fref{fig: gamma}.

Our understanding of this scenario is based on our analysis for $\gamma=\frac{\Gamma}{\sqrt{T}}$, where we show that the probability of $Q_T$ follows a large deviation principle
\begin{equation}
    \mathrm{P}\left(j=\frac{Q_T}{\sqrt{T}}\right)\asymp\e^{-\sqrt{T}\phi(j)}\label{current_dist_ldf}
\end{equation}
where $\phi(j)$ is the large deviation function (ldf) of current. Here, the symbol $\asymp$ implies that the ratio of logarithm of two sides in \eref{current_dist_ldf} converges to $1$ for large $T$. The ldf for $\Gamma\to \infty$ characterises the fast coupling regime. The $\Gamma\to 0$ gives the slow coupling regime where $\phi(j)$ vanishes and it reflects a different large deviations scaling.

Infinite one dimensional systems are known \cite{derrida_current_2009,Krapivsky2015} to exhibit a long-term memory of initial state and our system is not an exception. To emphasize this dependence mostly two kinds of initial states are considered. In one, the initial state is drawn from a stationary measure where atypical fluctuations in the initial state contribute to the current fluctuations. In the second, only typical configurations in the initial state are allowed. The long-time statistics of $Q_T$ or equivalently the ldf $\phi(j)$ is different in the two ensembles of initial states. Drawing analogy with disorder averages in spin-glass, these two initial states are referred as the annealed and the quenched ensembles, respectively. The analogy is evident for the cumulant generating function (cgf) in the two ensembles
\begin{equation}
    \fl\qquad\qquad\mu_{\mathcal{A}}(\lambda)=\log\left(\left<\left<e^{\lambda Q_T}\right>_{\mathrm{hist}}\right>_{\mathrm{init}}\right)\quad\mathrm{and}\quad\mu_{\mathcal{Q}}(\lambda)=\left<\log\left(\left<e^{\lambda Q_T}\right>_{\mathrm{hist}}\right)\right>_{\mathrm{init}}\label{anneal_quench_defn}
\end{equation}
where the subscripts ($\mathcal{A}$ and $\mathcal{Q}$) denote the respective (annealed and quenched) ensembles, the $\left<\right>_{\mathrm{hist}}$ denotes average over evolution, and $\left<\right>_{\mathrm{init}}$ denotes average over the initial state. For $\mu_{\mathcal{}Q}(\lambda)$ the initial-average of the slow-varying logarithm is dominated by the contribution from the most-typical configuration of the initial state. In the two examples we consider, the average density profile of the initial state is uniform at value $\rho_b$. 

\numparts
The essential physics can be seen already in the simplest example of the non-interacting particles. For the annealed ensemble
\begin{equation}
   \lim_{T\to\infty}\frac{\mu_{\mathcal{A}}(\lambda)}{\sqrt{T}}=\left[\rho_a\left(e^{\lambda}-1\right)+\rho_b\left(e^{-\lambda}-1\right)\right]\Omega_\Gamma\label{annealed_cgf}
\end{equation}
where the $\Gamma$ dependence in $\Omega_\Gamma$ is given by
\begin{equation}
    \Omega_\Gamma=\frac{2}{\sqrt{\pi}}-\frac{1-\e^{\Gamma^2}\mathrm{erfc}\,\Gamma}{\Gamma}.\label{E_func_boundary_rate}
\end{equation}
\endnumparts
The expression \eref{annealed_cgf} is similar to the cgf of current on an infinite line \cite{derrida_current_2009} and on a finite lattice coupled with two reservoirs \cite{Derrida2021}, with the difference in the overall factor $\Omega_\Gamma$.

\numparts
For the quenched ensemble
\begin{equation}
    \lim_{T\to\infty}\frac{\mu_{\mathcal{Q}}(\lambda)}{\sqrt{T}}=\rho_a\left(e^{\lambda}-1\right)\Omega_\Gamma+\rho_b\int_0^\infty\mathrm{d}x\,\log\left[1+\left(e^{-\lambda}-1\right)W_\Gamma(x)\right]\label{quenched_cgf}
\end{equation}
where
\begin{equation}
    W_\Gamma(x)=1-2\Gamma\int_0^\infty\mathrm{d}z\,\frac{\e^{-\pi^2z^2}}{\pi z\left(\Gamma^2+\pi^2z^2\right)}\left(\Gamma\sin\,\pi x z+\pi z\cos\,\pi x z\right).\label{W_func_boundary_rate}
\end{equation}
Note that $\int_0^\infty\mathrm{d}x\,W_\Gamma(x)=\Omega_\Gamma$. It is instructive to compare it to a similar result on the infinite line \cite{derrida_current_2009}. 
\endnumparts

\begin{figure}[tp]
    \quad\includegraphics[width=0.8\textwidth]{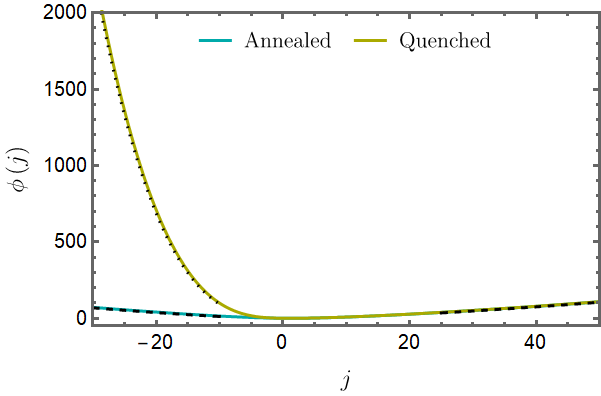}
    \caption{(color online) A comparison between the annealed and the quenched large deviation functions for the fast coupling limit ($\Gamma\to \infty$). The dashed lines indicate a $j\log j-j$ asymptotic whereas the dotted line indicates $j^3$ growth.}
    \label{fig: anneal vs quench strong / fast limit with fitting}
\end{figure}

The long-time scaling of the cgf implies the large deviation asymptotics in \eref{current_dist_ldf} where the ldf $\phi(j)$ is related to the cgf by a Legendre-Fenchel transformation 
\begin{equation}
    \mu(\lambda)\simeq \sqrt{T}\,\max_{j}\left(j\lambda-\phi(j)\right).\label{cgf_ldf_legendre}
\end{equation}
The ldf for the two ensembles for the fast coupling regime is shown in \fref{fig: anneal vs quench strong / fast limit with fitting}. For large positive $j$, the ldf for both ensembles grow as $j\log\frac{j}{\rho_a\Omega_\Gamma}-j$. For large negative $j$, the ldf for annealed ensemble grows as $\vert j\vert\log\frac{\vert j\vert}{\rho_b\Omega_\Gamma}-\vert j\vert$ whereas for the quenched it has a steeper growth of $\vert j\vert^3$. The positive tail of the quenched ldf is different for the infinite line \cite{derrida_current_2009}. 

For the SSEP our results are limited and we have expressions for only up to the third cumulant. The average of $Q_T$ is same for the two ensembles and is given by
\begin{equation}
    \frac{\left<Q_T\right>}{\sqrt{T}}\simeq\left(\rho_a-\rho_b\right)\Omega_\Gamma.\label{average_current_ssep}
\end{equation}
For the fast coupling limit, $\Omega_\Gamma\simeq\frac{2}{\sqrt{\pi}}$ and the resulting expression \eref{average_current_ssep} is consistent with the exact result in \cite{Tracy_2013}. 

The difference between the two ensembles emerge at the second and higher cumulants. Their expression is simpler in the fast coupling regime which we present below.  For annealed, the variance 
\begin{equation}
    \fl\qquad\frac{\left<Q_T^{\,2}\right>_\mathcal{A}}{\sqrt{T}}\simeq \frac{1}{\sqrt{\pi}}\left[2\left(\rho_a+\rho_b\right)-4\left(\sqrt{2}-1\right)\left(\rho_a^2+\rho_b^2\right)-4\left(3-2\sqrt{2}\right)\rho_a\rho_b\right] \label{eq:variance ann fast}
\end{equation}
whereas the variance for the quenched ensemble differs from the annealed by an additive term
\begin{equation}
    \frac{\left<Q_T^{\,2}\right>_\mathcal{Q}}{\sqrt{T}}\simeq\frac{\left<Q_T^{\,2}\right>_\mathcal{A}}{\sqrt{T}}- \frac{1}{\sqrt{\pi}}\bigg[2\left(2-\sqrt{2}\right)\rho_b\left(1-\rho_b\right)\bigg].\label{eq:variance qnch fast}
\end{equation}
The expression \eref{eq:variance ann fast} for the annealed ensemble is consistent with an earlier result \cite{Krapivsky2012} for $\rho_a=1$ and $\rho_b=0$. 

The third cumulant for the annealed is 
\begin{equation}
    \fl\frac{\left<Q_T^{\,3}\right>_\mathcal{A}}{\sqrt{T}}\simeq \frac{(\rho_a-\rho_b)}{\sqrt{\pi}} \!\!\left[2-\frac{8}{3} \left(9 \sqrt{2}-8 \sqrt{3}\right)\! (\rho_a-\rho_b)^2+12 \left(1- \sqrt{2}\right) \!(\rho_a+\rho_b-2 \rho_a \rho_b)\right]. \label{eq:Q3 ann fast}
\end{equation}
For the quenched we have an explicit result only for density $\rho_b=0$, where it is equal to the annealed result \eref{eq:Q3 ann fast}.

Our results for the two models are obtained using the hydrodynamic formalism of the MFT for the semi-infinite geometry with slow boundary. In recent years there is increasing interest in understanding the effect of slow coupling for the hydrodynamic description. To our knowledge, the earliest study in this direction was in \cite{Baldasso2017} for finite SSEP coupled with two reservoirs. The work established that the average macroscopic density profile in the steady state is a solution of the heat equation with Dirichlet, Robin, or Neumann boundary conditions, depending on the strength of the rates at the boundary. The work was later extended in \cite{Franco2019,Franco2016,Goncalves2018,Tsunoda2019,franceschini_hydrodynamical_2022,franceschini_symmetric_2021,Derrida2021}, particularly for small fluctuations in the steady state. Variants of the model with slow rates have also been studied in \cite{Bodineau2009,DeMasi2011,DeMasi2012,Franco2013,Redig2011,Landim2018,Erignoux2019}.

For a hydrodynamic approach to study fluctuations of $Q_T$ at large $T$, we define re-scaled coordinates $(x,t):=(\frac{i}{\sqrt{T}},\frac{\tau}{T})$ where $i=1, 2, \ldots $ is the site index of the semi-infinite lattice and $\tau$ is the microscopic time. We show that the average density $\rho_\mathrm{av}(\frac{i}{\sqrt{T}},\frac{\tau}{T})\simeq\left<n_i(\tau)\right>$ for large $T$, where $n_i(\tau)$ denotes the occupation of the $i$-th site at time $\tau$, follows the heat equation
\numparts\begin{equation}
    \frac{\partial\rho_{\mathrm{av}}(x,t)}{\partial t}=\frac{\partial^2\rho_{\mathrm{av}}(x,t)}{\partial x^2}\label{average_profile_diffusion_equation_transport_coefficient}
\end{equation}
with a Robin boundary condition
\begin{equation}
    \rho_{\mathrm{av}}(x,t)-\frac{1}{\Gamma}\frac{\partial\rho_{\mathrm{av}}(x,t)}{\partial x}=\rho_a\quad\textrm{at $x=0$}.\label{avg_density_boundary_condition}
\end{equation}\endnumparts
This is in agreement with rigorous mathematical results of \cite{Baldasso2017,Franco2019,Franco2016,Goncalves2018,Tsunoda2019}.

The density $\rho(\frac{i}{\sqrt{T}},\frac{\tau}{T})\simeq n_i(\tau)$ for large $T$ fluctuates around the average $\rho_{\mathrm{av}}(x,t)$. For SSEP, we argue that it satisfies a fluctuating hydrodynamics equation
\numparts\begin{equation}
    \frac{\partial\rho(x,t)}{\partial t}=
    \frac{\partial^2\rho(x,t)}{\partial x^2}+\frac{\partial\eta(x,t)}{\partial x}\label{eq:flhd full}
\end{equation}
with a fluctuating boundary condition
\begin{equation}
    -\frac{\partial\rho(x,t)}{\partial x}=\eta(x,t)+\xi(t)\quad \textrm{at $x=0$,}\label{eq:flhd boundary}
\end{equation}
where $\eta(x,t)$ is a weak Gaussian white noise with zero mean and covariance
\begin{equation}
    \left<\eta(x,t)\eta(x',t')\right>=\frac{1}{\sqrt{T}}\;2\rho(x,t)(1-\rho(x,t))\delta(x-x')\delta(t-t')
\end{equation}
and the noise $\xi(t)$ is delta correlated in time with a moment generating function
\begin{equation}
    \left<\e^{\sqrt{T}\int_0^1\mathrm{d}t\,h(t)\,\xi(t)}\right>\asymp \e^{\sqrt{T}\int_0^1\mathrm{d}t\,H_{\mathrm{bdry}}\left[\rho(0,t),\,h(t)\right]}
\end{equation}
for large $T$, with
\begin{equation}
    H_{\mathrm{bdry}}\left[\rho,h\right]=\Gamma\left[\rho_a\left(1-\rho\right)\left(\e^{h}-1\right)+\rho\left(1-\rho_a\right)\left(\e^{-h}-1\right)\right].\label{eq:Hbdry}
\end{equation}\endnumparts
The multiplicative noises in (\ref{eq:flhd full},\,\ref{eq:flhd boundary}) are interpreted with the It\^{o} convention. Our argument for (\ref{eq:flhd full},\,\ref{eq:flhd boundary}) is based on an Action formulation of the microscopic dynamics and a gradient expansion to obtain the hydrodynamic limit. Note that $\langle \xi(t) \rangle=\Gamma\left(\rho_a-\rho(0,t)\right)$ and with this \eref{avg_density_boundary_condition} is recovered.

We apply the MFT formulation for (\ref{eq:flhd full},\,\ref{eq:flhd boundary}) which gives the cgf of $Q_T$ as the solution of a variational problem. We explicitly solve the latter in the low density limit which corresponds to the non-interacting particles. The result is then verified using an exact solution of the microscopic dynamics. For arbitrary density, a perturbation solution of the variational problem gives our result for the first few cumulants of $Q_T$.

The rest of this article is organized as follows. In Section \ref{Sect: Macro}, we present the variational formalism for the cgf of $Q_T$. In Section \ref{sec:NI hydro} we solve the variational problem for the non-interacting particles and the solution is verified using microscopic dynamics in Section \ref{Sect: Micro}. Section \ref{sec:sep hydro} presents a perturbation solution of the variational problem for SSEP. We conclude with open directions in Section \ref{sec:concl}. In the \ref{Derive_Action_SSEP} we present an argument for (\ref{eq:flhd full},\,\ref{eq:flhd boundary}) and in \ref{app:derivation ave rho} we present a derivation for (\ref{average_profile_diffusion_equation_transport_coefficient},~\ref{avg_density_boundary_condition}).

\section{A fluctuating hydrodynamic approach for current fluctuations} \label{Sect: Macro}
We recall the variational formulation of the MFT \cite{bertini_macroscopic_2015,derrida_microscopic_2011,tailleur_mapping_2008} for current fluctuations in the SSEP on the semi-infinite line with slow coupling to reservoir. For large $T$, the time-integrated current across the system-reservoir boundary relates to the density
\begin{equation}
    Q_T=\sqrt{T}\int_0^{\infty}\mathrm{d}x\left(\rho(x,1)-\rho(x,0)\right)\label{time_int_current_defn}
\end{equation}
which is a consequence of particle number conservation in the bulk.
Within the MFT formulation the moment-generating function of $Q_T$ is expressed as a path integral over the density field $\rho(x,t)$ and a conjugate response field $\widehat{\rho}(x,t)$.
\begin{equation}
    \left<e^{\lambda Q_T}\right>_{\mathrm{hist}}=\int\mathcal{D}\left[\rho,\widehat{\rho}\right]\e^{-\sqrt{T}\;S\left[\rho,\widehat{\rho}\right]} \label{mgf_path_integral}
\end{equation}
where the action
\begin{eqnarray}
    \fl\qquad\qquad S\left[\rho,\widehat{\rho}\right]=&-\lambda \int_0^{\infty}\mathrm{d}x\left(\rho(x,1)-\rho(x,0)\right)+\mathcal{F}\left[\rho(x,0)\right]\nonumber\\ \fl\qquad\qquad&\;+\int_0^1\mathrm{d}t\left[\int_0^{\infty}\mathrm{d}x\left(\widehat{\rho}(x,t)\frac{\partial\rho(x,t)}{\partial t}\right)-H\left[\rho,\widehat{\rho}\right]\right]\label{action_hamilton_defn}
\end{eqnarray}
with an effective Hamiltonian \cite{tailleur_mapping_2008}
\begin{equation}
    \fl\; H[\rho,\widehat{\rho}]=H_{\mathrm{bdry}}\left[\rho(0,t),\widehat{\rho}(0,t)\right]+\int_0^{\infty}\mathrm{d}x\left(\frac{\sigma(\rho)}{2}\frac{\partial\widehat{\rho}(x,t)}{\partial x}-D(\rho)\frac{\partial\rho(x,t)}{\partial x}\right)\frac{\partial\widehat{\rho}(x,t)}{\partial x}\label{eq:Htotal}
\end{equation}
where the $H_{\mathrm{bdry}}$ defined in \eref{eq:Hbdry} represents the contribution from the slow coupling with the boundary reservoir. For the SSEP, the diffusivity $D(\rho)=1$ and the mobility $\sigma(\rho)=2\rho(1-\rho)$. We use the functions $D$ and $\sigma$ instead of their explicit expression to make the discussion generalizable for other models which differ in the transport coefficients and $H_\mathrm{bdry}$.  In \ref{Derive_Action_SSEP} we give a simplistic non-rigorous derivation of the action \eref{action_hamilton_defn} for the semi-infinite line. The expression \eref{action_hamilton_defn} for $\lambda=0$ and $\mathcal{F}=0$ is the Martin-Siggia-Rose-Janssen-De Dominicis action \cite{msrd_1,msrd_2,msrd_3,msrd_4} for the fluctuating hydrodynamics equation (\ref{eq:flhd full},\,\ref{eq:flhd boundary}) where $\widehat{\rho}$ is the response field.

The term $\mathcal{F}$ in \eref{action_hamilton_defn} is the contribution from the initial state, which for the annealed ensemble is the free energy of the initial equilibrium state of uniform average density $\rho_b$. For the semi-infinite line \cite{derrida_non-equilibrium_2007}
\begin{equation}
    \mathcal{F}\left[\rho(x,0)\right]=\int_0^{\infty}\mathrm{d}x\int_{\rho_b}^{\rho(x,0)}\mathrm{d}r\;\frac{2D(r)}{\sigma(r)}\left(\rho(x,0)-r\right)\label{initial_density_profile_ldf}
\end{equation}
where for convergence it is implicitly assumed that there are no fluctuations beyond a cutoff distance. 
For the quenched ensemble the leading contribution comes from $\rho(x,0)=\rho_b$ for which $\mathcal{F}\left[\rho_b\right]=0$.

For large $T$, the path-integral in \eref{mgf_path_integral} is dominated by the least-action path and the cgf \eref{anneal_quench_defn} is
\begin{equation}
    \mu(\lambda)\simeq -\sqrt{T}\min_{\rho\,,\,\widehat{\rho}}\,S\left[\rho,\widehat{\rho}\right]\label{cgf_min_action}
\end{equation}
with appropriate initial states.
We shall denote the least-action paths by $(\rho,\widehat{\rho})\equiv (q,p)$. The calculus of variations gives \cite{derrida_current_2009} the least-action paths as the solution of
\numparts\begin{eqnarray}
    \frac{\partial p(x,t)}{\partial t}=-D(q)\frac{\partial^2 p(x,t)}{\partial x^2}-\frac{\sigma'(q)}{2}{\left(\frac{\partial p(x,t)}{\partial x}\right)}^2\label{optimal_rho_hat_pde_transport}\quad\mathrm{and}\\    
    \frac{\partial q(x,t)}{\partial t}=\frac{\partial}{\partial x}\left(D(q)\frac{\partial q(x,t)}{\partial x}\right)-\frac{\partial}{\partial x}\left(\sigma(q)\frac{\partial p(x,t)}{\partial x}\right)\label{optimal_rho_pde_transport}
\end{eqnarray}\endnumparts
with the contribution from the $H_{\mathrm{bdry}}$ term included in the spatial boundary condition at $x=0$ for all $t$,
\numparts\begin{eqnarray}
   \fl\; D(q)\frac{\partial p}{\partial x}=-\frac{\delta H_{\mathrm{bdry}}}{\delta q(0,t)}=\Gamma\left[\rho_a\left(\e^{p(0,t)}-1\right)-\left(1-\rho_a\right)\left(\e^{-p(0,t)}-1\right)\right]\quad\mathrm{and}\label{boundary_systematic_first}\\
   \fl\;\sigma(q)\frac{\partial p}{\partial x}-D(q)\frac{\partial q}{\partial x}=\frac{\delta H_{\mathrm{bdry}}}{\delta p(0,t)}=\Gamma\left[\rho_a\left(1-q(0,t)\right)\e^{p(0,t)}-q(0,t)\left(1-\rho_a\right)\e^{-p(0,t)}\right].\label{boundary_systematic_second}
\end{eqnarray}\endnumparts
The optimization also gives \cite{derrida_current_2009} the temporal boundary conditions, which differs in the two ensembles. For the annealed ensemble, they are given as
\begin{equation}   p(x,0)=\lambda+\int_{\rho_b}^{q(x,0)}\mathrm{d}r\,\frac{2D(r)}{\sigma(r)}\quad\mathrm{and}\quad p(x,1)=\lambda. \label{initial_final_transport_coefficient}
\end{equation}
For the quenched ensemble, the temporal boundary conditions are
\begin{equation}
    q(x,0)=\rho_b\quad\mathrm{and}\quad p(x,1)=\lambda.\label{initial_density_quench}
\end{equation}

For both ensembles, using the least-action path \eref{optimal_rho_pde_transport} and the boundary condition \eref{boundary_systematic_second}, the expression \eref{action_hamilton_defn} for the minimal action \eref{cgf_min_action} simplifies  
\begin{eqnarray}
    \frac{\mu(\lambda)}{\sqrt{T}}\simeq&\lambda Q_T-\mathcal{F}\left[q(x,0)\right]-\int_0^1\mathrm{d}t\int_0^{\infty}\mathrm{d}x\,\frac{\sigma(q)}{2}{\left(\frac{\partial p}{\partial x}\right)}^2\nonumber\\
    &+\int_0^1\mathrm{d}t\left(H_{\mathrm{bdry}}-p(0,t)\frac{\delta H_{\mathrm{bdry}}}{\delta p(0,t)}\right).\label{eq:simpler mu}
\end{eqnarray}
The formal expression \eref{eq:simpler mu} for the cgf applies for diffusive systems with general transport coefficients and boundary Hamiltonian $H_{\mathrm{bdry}}$.

\begin{remark}
 In the fast coupling limit $\Gamma\to\infty$, (\ref{boundary_systematic_first},\,\ref{boundary_systematic_second}) give the Dirichlet boundary condition $q(0,t)=\rho_a$ and $p(0,t)=0$, which is often argued in the literature. In the slow coupling limit $\Gamma\to0$, we get the Neumann boundary condition $\partial_xq(x,t)=0=\partial_xp(x,t)$ at $x=0$.
\end{remark}
\begin{remark}
 For $\lambda=0$, the least-action-path is $q(x,t)=\rho_{\mathrm{av}}(x,t)$ and $p(x,t)=0$ which for SSEP is the solution of diffusion equation \eref{average_profile_diffusion_equation_transport_coefficient} with the Robin boundary condition \eref{avg_density_boundary_condition}.
\end{remark}

\begin{remark}
    There are obvious symmetries \cite{derrida_current_2009} evident from the action 
    \eref{action_hamilton_defn}. For the annealed ensemble, reflection symmetry implies $\mu_{\mathcal{A}}(\lambda\vert\rho_a,\rho_b)=\mu_{\mathcal{A}}(-\lambda\vert\rho_b,\rho_a)$ and particle-hole symmetry implies 
    $\mu_{\mathcal{A}}(\lambda\vert \rho_a,\rho_b)=\mu_{\mathcal{A}}(-\lambda\vert 1-\rho_a,1-\rho_b)$. There is an additional symmetry \cite{derrida_current_2009}
\begin{equation}
    \mu_{\mathcal{A}}(\lambda)=\mu_{\mathcal{A}}(f'(\rho_b)-f'(\rho_a)-\lambda)\label{eq:gc symm}
\end{equation}
due to invariance under time-reversal. The reflection symmetry and \eref{eq:gc symm} do not hold for the quenched ensemble.
\end{remark}

\section{Variational solution for non-interacting particles} \label{sec:NI hydro}
The variational problem for SSEP in \eref{cgf_min_action} is hard to solve in general. It is simpler to solve the low-density limit, which amounts to \cite{Derrida_2019,Rana_2023} setting $D(q)=1$ and $\sigma(q)=2q$ in \eref{action_hamilton_defn} with 
\begin{equation}
    H_{\mathrm{bdry}}\left[q(0,t),p(0,t)\right]=\Gamma\left[\rho_a\left(\e^{p(0,t)}-1\right)+q(0,t)\left(\e^{-p(0,t)}-1\right)\right].\label{boundary_action_nirw}
\end{equation}
The least-action path is (\ref{optimal_rho_pde_transport},\,\ref{optimal_rho_hat_pde_transport}) with $D(q)=1$ and $\sigma(q)=2q$ and the spatial boundary condition (\ref{boundary_systematic_first},\,\ref{boundary_systematic_second}) at $x=0$ becomes
\numparts\begin{eqnarray}
    \frac{\partial p(x,t)}{\partial x}=\Gamma\left(1-\e^{-p(0,t)}\right)\quad\mathrm{and}\label{boundary_first_nirw}\\
    2q(x,t)\frac{\partial p(x,t)}{\partial x}-\frac{\partial q(x,t)}{\partial x}=\Gamma\left(\rho_a\e^{p(0,t)}-q(0,t)\e^{-p(0,t)}\right).\label{boundary_second_nirw}
\end{eqnarray}\endnumparts

In order to simplify \eref{eq:simpler mu} for the non-interacting many particle system, we use an identity \cite{derrida_current_2009}
\begin{equation}
    q\left(\frac{\partial p}{\partial x}\right)^2=\frac{\partial}{\partial t}\left(qp\right)-\frac{\partial}{\partial x}\left(p\frac{\partial q}{\partial x}-q\frac{\partial p}{\partial x}-2qp\frac{\partial p}{\partial x}\right)
\end{equation}
extracted from the least-action path. With the help of the identity and the boundary condition (\ref{boundary_first_nirw},~\ref{boundary_second_nirw}) the cgf for the annealed ensemble in \eref{eq:simpler mu} simplifies to
\begin{equation}
    \lim_{T\to\infty}\frac{\mu_{\mathcal{A}}(\lambda)}{\sqrt{T}}=\Gamma\rho_a\int_0^1\mathrm{d}t\left(\e^{p(0,t)}-1\right)+\rho_b\int_0^\infty\mathrm{d}x\left(\e^{-\lambda}\e^{p(x,0)}-1\right),\label{eq:cgf ni formal anneal}
\end{equation}
while in the quenched ensemble, with $\mathcal{F}=0$ in \eref{eq:simpler mu}, the simplified cgf is
\begin{equation}
    \lim_{T\to\infty}\frac{\mu_{\mathcal{Q}}(\lambda)}{\sqrt{T}}=\Gamma\rho_a\int_0^1\mathrm{d}t\left(\e^{p(0,t)}-1\right)+\rho_b\int_0^\infty\mathrm{d}x\left(p(x,0)-\lambda\right).\label{eq:cgf ni formal quench}
\end{equation}

The linearized least-action equation for this problem is simple to solve using a known canonical transformation \cite{derrida_current_2009} $\left(G,R\right):=\left(\e^{p},q\e^{-p}\right)$. The new variables $G$ and $R$ are decoupled and follows
\begin{equation}
    \frac{\partial G(x,t)}{\partial t}=-\frac{\partial^2G(x,t)}{\partial x^2}\quad\mathrm{and}\quad\frac{\partial R(x,t)}{\partial t}=\frac{\partial^2R(x,t)}{\partial x^2}\label{anti_diffusion_G_diffusion_R}
\end{equation}
which are easy to solve using Green's function method.
In terms of new variables, the temporal boundary condition  \eref{initial_final_transport_coefficient} at final time $t=1$ gives
\begin{equation}
    G(x,1)=\e^\lambda\label{final_condition_G}
\end{equation}
while the initial conditions in (\ref{initial_final_transport_coefficient},~\ref{initial_density_quench}) impose different constraints
\begin{eqnarray}
    R(x,0)=\cases{\rho_b\e^{-\lambda} &for annealed ensemble,\\
    \frac{\rho_b}{G(x,0)} &for quenched ensemble.}\label{initial_condition_R}
\end{eqnarray}
For both the ensembles, the spatial boundary condition in (\ref{boundary_first_nirw},~\ref{boundary_second_nirw}) become,
\begin{equation}
    \fl\qquad G(x,t)-\frac{1}{\Gamma}\frac{\partial G(x,t)}{\partial x}=1\quad\mathrm{and}\quad R(x,t)-\frac{1}{\Gamma}\frac{\partial R(x,t)}{\partial x}=\rho_a\quad \textrm{at $x=0$.}\label{spatial_boundary_robin_G_R}
\end{equation}
The solution of the $G$-variable in either of the ensembles is
\begin{equation}
    G(x,t)=1+\left(\e^{\lambda}-1\right)\int_0^{\infty}\mathrm{d}y\,g(x,y,1-t)\label{G_solution_as_green}
\end{equation}
where
\begin{equation}
    \fl\quad g(x,y,t)=2\int_0^{\infty}\mathrm{d}z\; \frac{\e^{-\pi^2tz^2}}{\Gamma^2+\pi^2z^2}\left(\Gamma\sin\pi xz+\pi z\cos\pi xz\right)\left(\Gamma\sin\pi yz+\pi z\cos\pi yz\right)\label{green_function_arbitrary_boundary_rate}
\end{equation}
is the Green's function of the diffusion equation on the semi-infinite line $x>0$ for the Robin boundary condition $g-\Gamma^{-1}\partial_xg=0$ at $x=0$. (See details in \ref{green_function_details}.)

In a similar way, the solution for the $R$-variable is
\numparts\begin{eqnarray}
    \fl\qquad R(x,t)=\rho_a+\left(\rho_b\e^{-\lambda}-\rho_a\right)\int_0^{\infty}\textrm{d}y\,g(x,y,t)\qquad\quad\;\; \textrm{for annealed ensemble,}\label{R_anneal_solution_as_green}\\
    \fl\qquad R(x,t)=\rho_a+\int_0^{\infty}\mathrm{d}y\,g(x,y,t)\left(\frac{\rho_b}{G(y,0)}-\rho_a\right)\qquad\textrm{for quenched ensemble.}\label{R_quench_solution_as_green}
\end{eqnarray}\endnumparts

For the annealed ensemble using the solution for the least-action path in the expression for the cgf in \eref{eq:cgf ni formal anneal} we get
\begin{eqnarray}
    \frac{\mu_{\mathcal{A}}(\lambda)}{\sqrt{T}}\simeq&\Gamma\rho_a\left(\e^{\lambda}-1\right)\int_0^1\mathrm{d}t\int_0^{\infty}\mathrm{d}y\,g(0,y,1-t)\nonumber\\
    &+\rho_b\left(\e^{-\lambda}-1\right)\int_0^\infty\mathrm{d}x\left(1-\int_0^{\infty}\mathrm{d}y\,g(x,y,1)\right).
\end{eqnarray}
The last term in the equation above is re-expressed using an identity (see \ref{green_function_details} for a proof)
\begin{equation}
    \int_0^\infty\mathrm{d}x\left(1-\int_0^\infty\mathrm{d}y\,g(x,y,1)\right)=\Gamma\int_0^1\mathrm{d}t\int_0^\infty\mathrm{d}y\,g(0,y,1-t)\label{space_int_final_time_int_origin_green}
\end{equation}
that leads to the expression for cgf in \eref{annealed_cgf}.

For the quenched ensemble the cgf in \eref{eq:cgf ni formal quench} takes the form
\begin{eqnarray}
    \frac{\mu_{\mathcal{Q}}(\lambda)}{\sqrt{T}}\simeq&\Gamma\rho_a\left(\e^{\lambda}-1\right)\int_0^1\mathrm{d}t\int_0^{\infty}\mathrm{d}y\,g(0,y,1-t)\nonumber\\
    &+\rho_b\int_0^\infty\mathrm{d}x\log\left[1+\left(\e^{-\lambda}-1\right)\left(1-\int_0^{\infty}\mathrm{d}y\,g(x,y,1)\right)\right].
\end{eqnarray}
The two integrals in the first term gives the $\Omega_\Gamma$ in \eref{E_func_boundary_rate}, which leads to expression reported in \eref{quenched_cgf}.
\begin{figure}
    \quad\includegraphics[width=0.8\textwidth]{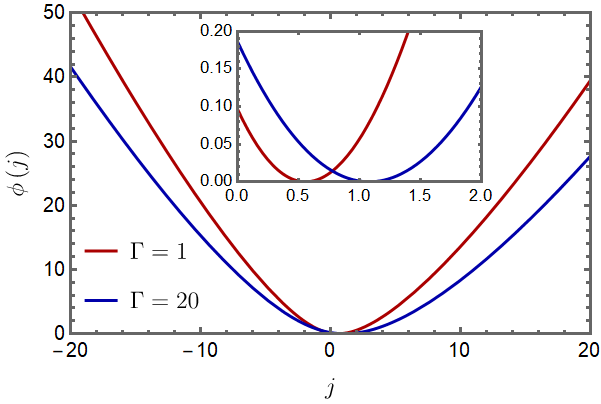}
    \caption{(color online) The large deviation function $\phi(q)$ for the annealed ensemble with $\rho_a=2$ and $\rho_b=1$ for two different values of $\Gamma$ indicated in the \textit{legend}.  The inset shows a zoom-in plot near the minimum of the ldf to indicate the dependence of the average current on $\Gamma$.}
    \label{fig: annealed ldf different boundary rate}
\end{figure}

\subsection{Large deviations function}\label{sect:ldf_nirw}
The ldf $\phi(j)$ relates to the cgf by the Legendre-Fenchel transformation \eref{cgf_ldf_legendre}. For the annealed case,  using \eref{annealed_cgf} we get an explicit expression
\begin{equation}
    \fl\qquad\phi_\mathcal{A}(j)=j\log\frac{j+\sqrt{j^2+4\rho_a\rho_b\Omega_\Gamma^{\,2}}}{2\rho_a\Omega_\Gamma}-\frac{4\rho_a\rho_b\Omega_\Gamma^{\,2}}{j+\sqrt{j^2+4\rho_a\rho_b\Omega_\Gamma^{\,2}}}-j+\left(\rho_a+\rho_b\right)\Omega_\Gamma
\end{equation}
where, $\Omega_\Gamma$ is the boundary rate dependent factor defined in \eref{E_func_boundary_rate}. The ldf is plotted in \fref{fig: annealed ldf different boundary rate} which illustrates the dependence on $\Gamma$.

For the quenched ensemble from \eref{quenched_cgf}, we obtain a parametric expression
\numparts\begin{eqnarray}
    \phi_\mathcal{Q}(j)=&\lambda\rho_a\e^{\lambda}\Omega_\Gamma-\lambda\rho_b\e^{-\lambda}\int_0^{\infty}dx\,\frac{W_\Gamma(x)}{1+\left(\e^{-\lambda}-1\right)W_\Gamma(x)}\nonumber\\
    &-\rho_a\left(\e^{\lambda}-1\right)\Omega_\Gamma-\rho_b\int_0^\infty dx\,\log\left[1+\left(\e^{-\lambda}-1\right)W_\Gamma(x)\right]
\end{eqnarray}
with the current $j$ depending on the $\lambda$-parameter via
\begin{equation}
    j=\rho_a\e^{\lambda}\Omega_\Gamma-\rho_b\e^{-\lambda}\int_0^{\infty}dx\,\frac{W_\Gamma(x)}{1+\left(\e^{-\lambda}-1\right)W_\Gamma(x)}
\end{equation}\endnumparts
where $W_\Gamma$ is defined in \eref{W_func_boundary_rate}.

\begin{figure}
     \quad\includegraphics[width=0.8\textwidth]{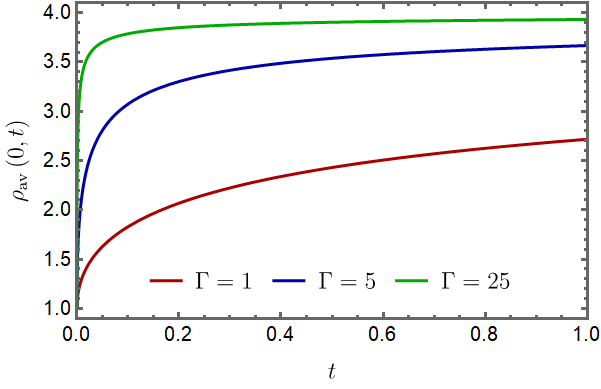}
     \caption{(color online) Time evolution of the average density \eref{eq:rhoave sol} near the reservoir for different boundary rates $\Gamma$ indicated in the \textit{legend}. We have set $\rho_a=4$ and $\rho_b=1$.}
     \label{fig: evolution of average density at boundary}
 \end{figure}

\begin{remark}
The least-action path $q(x,t)=G(x,t)R(x,t)$ with the solution (\ref{G_solution_as_green},~\ref{R_anneal_solution_as_green},~\ref{R_quench_solution_as_green}) gives the most-probable density evolution contributing to an atypical value of $Q_T=\mu^\prime(\lambda)$ for large $T$ in the respective ensemble. For $\lambda=0$ the $q(x,t)$ is the average density $\rho_{\mathrm{av}}(x,t)$ which is identical in both ensembles and evolves as
\begin{equation}
    \fl\qquad\rho_{\mathrm{av}}(x,t)=\rho_a+2\Gamma\left(\rho_b-\rho_a\right)\int_0^\infty\mathrm{d}z\,\frac{\e^{-\pi^2tz^2}}{\pi z\left(\Gamma^2+\pi^2z^2\right)}\left(\Gamma\sin\pi xz+\pi z\cos\pi xz\right).\label{eq:rhoave sol}
\end{equation}
At the boundary the density $\rho_{\mathrm{av}}(0,t)$ is different from the reservoir density $\rho_a$ which it asymptotically approaches as $\rho_a+\frac{\rho_b-\rho_a}{\Gamma\sqrt{{\pi}t}}$ for large $t$. A plot of the evolution is shown in  \fref{fig: evolution of average density at boundary}.
\end{remark}

\section{An exact microscopic analysis} \label{Sect: Micro}
The derivation we followed (\ref{Derive_Action_SSEP}) for the hydrodynamic formulation (\ref{mgf_path_integral}) rests on certain uncontrolled approximations to get the hydrodynamic limit. It is therefore important to verify the results for the non-interacting particles in Sec.~\ref{sec:NI hydro} using an independent exact analysis of the microscopic dynamics. 

Similar to the SSEP in \fref{fig: SSEP Slow}, for the non-interacting dynamics the nearest neighbour jump rates in the bulk sites are chosen as $1$ where particles move without an exclusion interaction. The boundary site is coupled with the reservoir of density $\rho_a$ which is modelled by the injection rate $\gamma \rho_a$ from the reservoir and the extraction rate $\gamma$ to the reservoir. The site index of the lattice are denoted by positive integers $i=1,2,3,\ldots$ and the microscopic time coordinates by $0\le\tau\le T$. The occupancy of the $i$-th lattice site in the initial configuration is denoted by $n_i$ which we draw from a Poisson distribution $P(n_i)=\frac{\rho_b^{n_i}\, e^{-\rho_b}}{n_i!}$ with uniform average occupation $\left<n_i\right>=\rho_b$.

Presence of the boundary makes the calculation slightly different from the analysis \cite{derrida_current_2009,sadhu_large_2015} on an infinite line. 
If a particle is in the $i$-th lattice site, then either it exits the lattice into the reservoir by time $\tau$ or it continues to be somewhere in the lattice till time $\tau$. We denote the probability of such events by $\mathcal{E}_i(\tau)$ (escape probability), and $\mathcal{S}_i(\tau)$ (survival probability), respectively. Bearing in mind that the particles are non-interacting, we write the joint statistics in terms of single-particle statistics.
\begin{eqnarray}
    \fl\qquad \left<e^{\lambda Q_T}\right>_{\mathrm{hist}}=&\prod_{k=0}^{T/\mathrm{d}\tau}\left[1-\gamma \rho_a\,\mathrm{d}\tau\,\mathcal{S}_1(T-k\mathrm{d}\tau)+e^{\lambda}\gamma \rho_a\,\mathrm{d}\tau\,\mathcal{S}_1(T-k\mathrm{d}\tau)\right]\nonumber\\
    \fl&\times\prod_{i=1}^{\infty}\left[1-\mathcal{E}_i(T)+e^{-\lambda}\mathcal{E}_i(T)\right]^{n_i}\label{e_lambda_QT_average_history}
\end{eqnarray}
where we have discretized the total time duration $T$ into infinitesimal intervals of length $\mathrm{d}\tau$. The first product is the contribution from particles that are injected from the reservoir into the lattice during the time interval $[0,T]$ and the second product is the contribution from particles present in the initial state. In the vanishing $\mathrm{d}\tau$ limit,
\begin{equation}
    \fl\qquad\left<e^{\lambda Q_T}\right>_{\mathrm{hist}}=\exp\left[\gamma\rho_a\left(e^{\lambda}-1\right)\int_0^T\mathrm{d}\tau\,\mathcal{S}_1(\tau)\right]\prod_{i=1}^{\infty}\left[1+\left(e^{-\lambda}-1\right)\mathcal{E}_i(T)\right]^{n_i}.\label{e_lambda_QT_average_history_second}
\end{equation}

The escape probabilities $\mathcal{E}_i(\tau)$ are solution of the rate equation
\begin{equation}
    \frac{\mathrm{d}\mathcal{E}_i(\tau)}{\mathrm{d}\tau}=\cases{-\left(1+\gamma\right)\mathcal{E}_1(\tau)+\mathcal{E}_2(\tau)+\gamma&for $i=1$,\\
    \mathcal{E}_{i-1}(\tau)-2\mathcal{E}_i(\tau)+\mathcal{E}_{i+1}(\tau)&for $i\geq2$\\}\label{escape_probability_time_evolution_pde}
\end{equation}
with the initial condition $\mathcal{E}_i(0)=0$. The survival probability simply relates to the escape probability $\mathcal{E}_i(\tau)+\mathcal{S}_i(\tau)=1$. Using this relation and \eref{escape_probability_time_evolution_pde} we find an identity
\begin{equation}
    \gamma\int_0^T\mathrm{d}\tau\,\mathcal{S}_1(\tau)=\sum_{i=1}^{\infty}\mathcal{E}_i(T)\label{reln_bw_escape_survival_time_escape_site}
\end{equation}
that will be used shortly.

For an explicit expression of $\mathcal{E}_i(\tau)$ we solve  \eref{escape_probability_time_evolution_pde} using the Laplace transformation
\begin{equation}
    \widehat{\mathcal{E}}_i(s)\equiv\mathcal{L}\left[\mathcal{E}_i\right](s):=\int_0^\infty\mathrm{d}\tau\;e^{-s\tau}\mathcal{E}_i(\tau)\label{escape_laplace_transf_defn}
\end{equation}
which follows 
\begin{equation}
    s\,\widehat{\mathcal{E}}_i(s)=\cases{\widehat{\mathcal{E}}_2(s)-\left(1+\gamma\right)\widehat{\mathcal{E}}_1(s)+\frac{\gamma}{s}&for $i=1$,\\
    \widehat{\mathcal{E}}_{i+1}(s)-2\widehat{\mathcal{E}}_i(s)+\widehat{\mathcal{E}}_{i-1}(s)&for $i\geq2$.\\}\label{escape_probability_time_evolution_pde_Laplace}
\end{equation}
The solution that vanishes at $i\to \infty$ is
\begin{equation}
    \widehat{\mathcal{E}}_i(s)=\frac{\gamma}{s}{\left(\gamma+\frac{s}{2}+\sqrt{s+\frac{s^2}{4}}\right)}^{-1}{\left(1+\frac{s}{2}-\sqrt{s+\frac{s^2}{4}}\right)}^{i-1}.\label{escape_probability_laplace_solution}
\end{equation}
Inverting the Laplace transformation \eref{escape_laplace_transf_defn} to get $\mathcal{E}_i(\tau)$ is hard. However, for the large $\tau$ limit, asymptotic expression for $\mathcal{E}_i(\tau)$ can be obtained which is presented in the next section.

\subsection{Annealed ensemble}
Averaging \eref{e_lambda_QT_average_history_second} over the initial configurations which are drawn from a Poisson distribution with an average $\rho_b$, we get
\begin{eqnarray}
    \left<\left<e^{\lambda Q_T}\right>_{\mathrm{hist}}\right>_{\mathrm{init}}=&\exp\left[\gamma\rho_a\left(e^{\lambda}-1\right)\int_0^T\mathrm{d}\tau\,\mathcal{S}_1(\tau)\right]\nonumber\\
    &\times\sum_{n_i=0}^{\infty}\prod_{i=1}^{\infty}\left\{{\left[1+\left(e^{-\lambda}-1\right)\mathcal{E}_i(T)\right]}^{n_i}\frac{\rho_b^{n_i}e^{-\rho_b}}{n_i!}\right\}
\end{eqnarray}
which after simplifying and using \eref{reln_bw_escape_survival_time_escape_site} gives the cgf in \eref{anneal_quench_defn} as
\begin{equation}
    \mu_{\mathcal{A}}(\lambda)=\gamma\left[\rho_a\left(e^{\lambda}-1\right)+\rho_b\left(e^{-\lambda}-1\right)\right]\int_0^T\mathrm{d}\tau\,\mathcal{S}_1(\tau).\label{eq:muA intermediate expr}
\end{equation}

To compute the integral in the above expression we use $\mathcal{S}_1(\tau)=1-\mathcal{E}_1(\tau)$ and write the Laplace transformation 
\begin{equation}
\mathcal{L}\left[\int_0^T\mathrm{d}\tau\,\mathcal{S}_1(\tau)\right](s)=\frac{1-s\, \widehat{\mathcal{E}}_1(s)}{s^2}\simeq \frac{\sqrt{s}}{s^2\left(\gamma+\sqrt{s}\right)}
\end{equation}
for small $s$ where we used \eref{escape_probability_laplace_solution}. In the large $T$ limit the leading contribution to the integral comes from small $s$. By taking the inverse Laplace transformation we get  
\begin{equation}
    \int_0^T\mathrm{d}\tau\,\mathcal{S}_1(\tau)\simeq\frac{\sqrt{T}}{\gamma}\left(\frac{2}{\sqrt{\pi}}-\frac{1-\e^{\gamma^2T}\mathrm{erfc}\,\gamma\sqrt{T}}{\gamma\sqrt{T}}\right)\label{sum_sites_escape}
\end{equation}
for large $T$. Using this asymptotic in \eref{eq:muA intermediate expr} with $\gamma=\frac{\Gamma}{\sqrt{T}}$ gives the expression \eref{annealed_cgf}.

\subsection{Quenched ensemble}
Taking the logarithm of \eref{e_lambda_QT_average_history_second} and then averaging over the initial configurations we get
\begin{eqnarray}
    \fl\qquad\left<\log\left[\left<e^{\lambda Q_T}\right>_{\mathrm{hist}}\right]\right>_{\mathrm{init}}=&\gamma\rho_a\left(e^{\lambda}-1\right)\int_0^T\mathrm{d}\tau\,\mathcal{S}_1(\tau)\nonumber\\
    \fl&+\sum_{n_i=0}^\infty\sum_{i=1}^\infty\left\{\log{\left[1+\left(e^{-\lambda}-1\right)\mathcal{E}_i(T)\right]}^{n_i}\frac{\rho_b^{n_i}e^{-\rho_b}}{n_i!}\right\}.
\end{eqnarray}
This gives the cgf \eref{anneal_quench_defn} in the quenched ensemble
\begin{equation}
    \mu_{\mathcal{Q}}(\lambda)=\gamma\rho_a\left(e^{\lambda}-1\right)\int_0^T\mathrm{d}\tau\,\mathcal{S}_1(\tau)+\rho_b\sum_{i=1}^\infty\log\left[1+\left(e^{-\lambda}-1\right)\mathcal{E}_i(T)\right].\label{eq:muQ micro NI}
\end{equation}

To obtain the leading asymptotic for large $T$ in the parameter regime of our interest we set $\gamma=\frac{\Gamma}{\sqrt{T}}$. Asymptotic of the first term in \eref{eq:muQ micro NI} is given by \eref{sum_sites_escape} whereas for the second term we take the diffusive scaling $\mathcal{E}_i(\tau)\simeq \mathcal{E}(\frac{i}{\sqrt{T}},\frac{\tau}{T})$ and write
\begin{equation}
    \frac{\mu_{\mathcal{Q}}(\lambda)}{\sqrt{T}}=\rho_a\left(e^{\lambda}-1\right)\Omega_\Gamma+\rho_b\int_0^\infty\mathrm{d}x\,\log\left[1+\left(e^{-\lambda}-1\right)\mathcal{E}(x,1)\right]\label{eq:mu Q intr}
\end{equation}
with $\Omega_\Gamma$ defined in \eref{E_func_boundary_rate}. The scaling of $\mathcal{E}_i(\tau)$ implies its Laplace transform $\widehat{\mathcal{E}}_i(s)\simeq T\,\widehat{\mathcal{E}}(\frac{i}{\sqrt{T}},s T)$ for large $T$ with 
\begin{equation}
    \widehat{\mathcal{E}}(x,r)=\frac{\Gamma }{r\left(\Gamma+\sqrt{r}\right)}e^{-x\sqrt{r}}\label{eq:e hat scaling}
\end{equation}
obtained from \eref{escape_probability_laplace_solution} with $\gamma=\frac{\Gamma}{\sqrt{T}}$. The inverse Laplace transformation of \eref{eq:e hat scaling} gives $\mathcal{E}(x,t)=W_{\Gamma\sqrt{t}}\left(\frac{x}{\sqrt{t}}\right)$ in \eref{W_func_boundary_rate} which substituted in \eref{eq:mu Q intr} confirms the cgf \eref{quenched_cgf}.

\section{Perturbation solution for the Symmetric Simple Exclusion Process}\label{sec:sep hydro}
The variational problem \eref{cgf_min_action} for SSEP is hard to solve explicitly. Corresponding variational problem on the infinite line \cite{derrida_current_2009} has been recently solved \cite{mallick2022exact} using a remarkable connection to an integrable model. While the difference between the variational problems for the infinite and semi-infinite geometry is only in the spatial boundary condition,  the later is hard to solve at present. 

In this article, we use a straightforward perturbation expansion in small $\lambda$ to solve the variational problem order by order. This gives a systematic approach for the cumulants of the current. Similar perturbation expansion has been used for the statistics of current \cite{krapivsky_fluctuations_2012} and activity \cite{KrapivskyMelting2015} on an infinite line and tracer statistics in single file diffusion \cite{Krapivsky2015,Rana_2023}.

For $\lambda=0$, the least-action solution for either ensembles is $p(x,t)=0$ and $q(x,t)=\rho_\mathrm{av}(x,t)$, where $\rho_\mathrm{av}(x,t)$ is given in \eref{eq:rhoave sol} which follows from (\ref{average_profile_diffusion_equation_transport_coefficient},~\ref{avg_density_boundary_condition}). For small $\lambda$ we write a perturbation expansion
\numparts\begin{eqnarray}
    q=\rho_\mathrm{av}+\lambda q_1+\lambda^2q_2+\ldots,\\
    p=\lambda p_1+\lambda^2p_2+\ldots. \label{eq:perturbation of fields}
\end{eqnarray}\endnumparts
Substituting the expansion in (\ref{optimal_rho_pde_transport},~\ref{optimal_rho_hat_pde_transport}) gives the least-action path at the linear order  
\numparts
\begin{eqnarray}
    \frac{\partial p_1(x,t)}{\partial t}=-\frac{\partial^2p_1(x,t)}{\partial x^2},\label{eq:eq for p1}\\
    \frac{\partial q_1(x,t)}{\partial t}=\frac{\partial^2q_1(x,t)}{\partial x^2}-\frac{\partial}{\partial x}\left(\sigma(\rho_{\mathrm{av}})\frac{\partial p_1(x,t)}{\partial x}\right),\label{eq:eq for q1}
    \end{eqnarray}\endnumparts
and at the quadratic order
\numparts\begin{eqnarray}
    \fl\qquad\quad\frac{\partial p_2(x,t)}{\partial t}=-\frac{\partial^2p_2(x,t)}{\partial x^2}-\frac{\sigma'(\rho_{\mathrm{av}})}{2}{\left(\frac{\partial p_1(x,t)}{\partial x}\right)}^2,\label{eq:eq for p2}\\
    \fl\qquad\quad\frac{\partial q_2(x,t)}{\partial t}=\frac{\partial^2q_2(x,t)}{\partial x^2}-\frac{\partial}{\partial x}\left(q_1(x,t)\sigma'(\rho_{\mathrm{av}})\frac{\partial p_1(x,t)}{\partial x}+\sigma(\rho_{\mathrm{av}})\frac{\partial p_2(x,t)}{\partial x}\right).\label{eq:eq for q2}
\end{eqnarray}\endnumparts

The Robin boundary conditions (\ref{boundary_systematic_first},~\ref{boundary_systematic_second}) gives the spatial boundary condition for these perturbation fields at $x=0$. At the linear order, we obtain
\numparts
\begin{eqnarray}
    \fl\qquad p_1(0,t)-\left.\frac{1}{\Gamma}\frac{\partial p_1(x,t)}{\partial x}\right\vert_{x=0}=0,\label{spatial_p1_0}\\
    \fl\qquad q_1(0,t)-\left.\frac{1}{\Gamma}\frac{\partial q_1(x,t)}{\partial x}\right\vert_{x=0}=p_1(0,t)\left(\rho_a-\rho_{\mathrm{av}}(0,t)\right)\left(1-2\rho_{\mathrm{av}}(0,t)\right)\equiv C_1(t),\label{spatial_q1_C1}
\end{eqnarray}\endnumparts
while at the quadratic order, we obtain
\numparts
\begin{eqnarray}
    \fl\qquad p_2(0,t)-\left.\frac{1}{\Gamma}\frac{\partial p_2(x,t)}{\partial x}\right\vert_{x=0}=&\frac{p_1^2(0,t)\left(1-2\rho_a\right)}{2}\equiv C_2(t),\label{spatial_p2_C2}\\
    \fl\qquad q_2(0,t)-\left.\frac{1}{\Gamma}\frac{\partial q_2(x,t)}{\partial x}\right\vert_{x=0}=&\frac{p_1^2(0,t)\left(\rho_a-\rho_{\mathrm{av}}(0,t)\right)}{2}\nonumber\\
    \fl\qquad&-p_1(0,t)q_1(0,t)\left(1+2\rho_a-4\rho_{\mathrm{av}}(0,t)\right)\nonumber\\
    \fl\qquad&+p_1^2(0,t)\left(1-2\rho_a\right)\rho_{\mathrm{av}}(0,t)\left(1-\rho_{\mathrm{av}}(0,t)\right)\nonumber\\
    \fl\qquad&+p_2(0,t)\left(\rho_a-\rho_{\mathrm{av}}(0,t)\right)\left(1-2\rho_{\mathrm{av}}(0,t)\right)\equiv C_3(t).\label{spatial_q2_C3}
\end{eqnarray}
\endnumparts

In the perturbation expansion, the fields at certain order in $\lambda$ only depends on the lower-order fields and therefore they can be systematically solved order-by-order. Using the perturbation expansion \eref{eq:perturbation of fields} in the expression \eref{eq:simpler mu} gives a series solution the cgf 
\begin{equation}
    \mu(\lambda)=\lambda \langle Q_T\rangle +\frac{\lambda^2}{2!}\langle Q_T^2\rangle + \frac{\lambda^3}{3!}\langle Q_T^3\rangle + \cdots
\end{equation}
where the coefficients in each order are the cumulants of the current. A similar hierarchical dependence prevails. Namely, the first cumulant $\langle Q_T\rangle$ is expressed in terms of the leading order field $\rho_{\mathrm{av}}$, the second cumulant $\langle Q_T^2\rangle$ in terms of $(q_1,p_1)$, and so on. 

For the quenched ensemble, the expression for the average is
\begin{equation}
    \frac{\left<Q_T\right>_{\mathcal{Q}}}{\sqrt{T}}\simeq-\int_0^1\mathrm{d}t\left.\frac{\partial\rho_{\mathrm{av}}(x,t)}{\partial x}\right|_{x=0},\label{avg_current_slow_general}
\end{equation}
where the average density $\rho_{\mathrm{av}}(x,t)=\rho_a+\left(\rho_b-\rho_a\right)\int_0^\infty\mathrm{d}y\,g(x,y,t)$ with the Green's function in \eref{green_function_arbitrary_boundary_rate}. The expression in \eref{avg_current_slow_general} can be explicitly evaluated and we arrive at the result \eref{average_current_ssep}. Note that this is identical to the expression we obtained for the non-interacting particles.

Similarly from the successive orders in the perturbation series we obtain the variance
\numparts\begin{eqnarray}
    \fl \frac{\left<Q_T^{\,2}\right>_{\mathcal{Q}}}{\sqrt{T}}\simeq\int_0^1\mathrm{d}t\int_0^{\infty}\mathrm{d}x\,\sigma(\rho_{\mathrm{av}}(x,t))\left(\frac{\partial p_1(x,t)}{\partial x}\right)^2+\Gamma\int_0^1\mathrm{d}t\,B_2\left[\rho_{\mathrm{av}}(0,t),p_1(0,t)\right]\label{eq:variance quench formal}
\end{eqnarray}
with $\sigma(\rho)=2\rho(1-\rho)$ and
\begin{equation}
    B_2\left[\rho_{\mathrm{av}},p_1\right]=p_1^2\left(2\rho_{\mathrm{av}}\rho_a-\rho_{\mathrm{av}}-\rho_a\right),\label{variance boundary term}
\end{equation}\endnumparts
and the third cumulant
\numparts\begin{eqnarray}
    \frac{\left<Q_T^{\,3}\right>_{\mathcal{Q}}}{\sqrt{T}}\simeq&3\int_0^1\mathrm{d}t\int_0^{\infty}\mathrm{d}x\,\left[q_1(x,t)\sigma'(\rho_{\mathrm{av}}(x,t))\left(\frac{\partial p_1(x,t)}{\partial x}\right)^2\right]\nonumber\\
    &+\Gamma\int_0^1\mathrm{d}t\,B_3\left[\rho_{\mathrm{av}}(0,t),p_1(0,t),q_1(0,t),p_2(0,t)\right]\label{eq:third cumulant quench formal}
\end{eqnarray}
with 
\begin{eqnarray}
    B_3\left[\rho_{\mathrm{av}},p_1,q_1,p_2\right]=p_1\Big(&2\rho_{\mathrm{av}}p_1^2+12\rho_{\mathrm{av}}p_2\rho_a-6\rho_{\mathrm{av}}p_2-2p_1^2\rho_a\nonumber\\
    &+6q_1p_1\rho_a-3q_1p_1-6p_2\rho_a\Big),\label{third cumulant boundary term}
\end{eqnarray}\endnumparts 
and so on.

Similar analysis applies for the annealed ensemble. The average current is same as given in \eref{average_current_ssep} for the quenched ensemble. The expression for the higher cumulants is however different and they are presented in the \sref{sec:annealed perturbation}.

\subsection{Quenched ensemble}
The temporal boundary condition \eref{initial_density_quench} gives the condition for the perturbation fields
\numparts\begin{eqnarray}
    p_1(x,1)=1,\;p_2(x,1)=0,\;\ldots,\label{final_p_que_ann}\\
    q_1(x,0)=0,\;q_2(x,0)=0,\;\ldots.\label{initial_q_quench}
\end{eqnarray}\endnumparts

The corresponding field equations (\ref{eq:eq for p1},\;\ref{eq:eq for q1}) and (\ref{eq:eq for p2},\;\ref{eq:eq for q2}) can be solved order by order for arbitrary $\Gamma$, first solving for $p_k$  at a perturbation order $k$ in terms of the Green's function $g(x,y,t)$ in  \eref{green_function_arbitrary_boundary_rate} and then $q_k$ in terms of the Green's function $\widehat{g}(x,y,t)$ defined in \eref{eq:modified greens sol} for a different boundary condition. However, their expression for arbitrary $\Gamma$ are cumbersome and we do not present those results in this article. In comparison the fast coupling limit ($\Gamma\to\infty$) yields simpler expressions and captures the interesting non-equilibrium regime. We present rest of the analysis for this fast coupling limit where the boundary conditions (\ref{spatial_p1_0},\,\ref{spatial_q2_C3}) on the perturbation fields reduce to Dirichlet conditions. 

The mean \eref{avg_current_slow_general} in the fast coupling limit yields
\begin{equation}
    \frac{\left<Q_T\right>}{\sqrt{T}}\simeq \frac{2}{\sqrt{\pi}}\left(\rho_a-\rho_b\right)
\end{equation}
which is the $\Gamma\to \infty$ limit of \eref{average_current_ssep}. The variance \eref{eq:variance quench formal} in the fast coupling limit gives
\begin{equation}
    \frac{\left<Q_T^{\,2}\right>_{\mathcal{Q}}}{\sqrt{T}}\simeq\frac{1}{\pi}\int_0^1\mathrm{d}t\int_0^\infty\mathrm{d}x\,\sigma(\rho_{\mathrm{av}}(x,t))\left(\frac{\e^{-x^2/2\left(1-t\right)}}{1-t}\right)\label{eq:variance quench not explicit}
\end{equation}
with $\rho_{\mathrm{av}}(x,t)=\rho_a+\left(\rho_b-\rho_a\right)\mathrm{erf}\left(\frac{x}{2\sqrt{t}}\right)$ which leads to the explicit result \eref{eq:variance qnch fast}. The third cumulant \eref{eq:third cumulant quench formal} in the fast coupling limit gives
\begin{equation}
    \frac{\left<Q_T^{\,3}\right>_{\mathcal{Q}}}{\sqrt{T}}\simeq -\frac{6}{\pi}\int_0^1\mathrm{d}t\int_0^{\infty}\mathrm{d}x\,\sigma'(\rho_{\mathrm{av}}(x,t))\partial_x\psi(x,t)\left(\frac{\e^{-x^2/2\left(1-t\right)}}{1-t}\right) \label{eq:Q3 quench not explicit}
\end{equation}
where
\begin{eqnarray}
    \fl\psi(x,t)=\int_0^t\mathrm{d}s\int_0^\infty\mathrm{d}y\,\sigma(\rho_{\mathrm{av}}(y,s))\frac{\e^{-y^2/4\left(1-s\right)}}{\sqrt{4\pi\left(1-s\right)}}\left(\frac{\e^{-{\left(y-x\right)}^2/4(t-s)}+\e^{-{\left(y+x\right)}^2/4(t-s)}}{\sqrt{4\pi\left(t-s\right)}}\right)\label{eq:psi1 tilde}
\end{eqnarray}
At present have not succeeded in explicitly completing the integrals in \eref{eq:Q3 quench not explicit}. For $\rho_b=0$, where quenched and annealed ensembles are identical we have verified that \eref{eq:Q3 quench not explicit} is consistent with the explicit result \eref{eq:Q3 ann fast}.

\begin{remark}
    The solutions (\ref{eq:variance quench not explicit},\,\ref{eq:Q3 quench not explicit}) formally extend for systems with unit diffusivity $D(\rho)=1$ and general mobility $\sigma(\rho)$.
\end{remark}

\subsection{Annealed ensemble}\label{sec:annealed perturbation}
The temporal boundary conditions \eref{initial_final_transport_coefficient} for $p(x,1)$ are identical for the two ensembles, which leads to the same solution for the $p_1(x,t)$ and $p_2(x,t)$ as in the quenched ensemble.

For $q(x,t)$ the boundary condition \eref{initial_final_transport_coefficient} at $t=0$ gives
\numparts\begin{eqnarray}\label{initial_q_anneal}
    q_1(x,0)=\frac{\sigma(\rho_b)}{2}\left(p_1(x,0)-1\right),\\
    q_2(x,0)=\frac{\sigma(\rho_b)p_2(x,0)}{2}+\frac{\sigma'(\rho_b)\sigma(\rho_b)}{8}\left(p_1(x,0)-1\right)^2,
\end{eqnarray}\endnumparts
which can be explicitly written using the solution
\numparts\begin{eqnarray}
    \fl\qquad p_1(x,0)=\mathrm{erf}\left(\frac{x}{2}\right),\\
    \fl\qquad p_2(x,0)=\int_0^1\mathrm{d}t\int_0^\infty\mathrm{d}y\,\sigma'(\rho_{\mathrm{av}}(y,1-t))\frac{\e^{-y^2/2t}}{2\pi t}\left(\frac{\e^{-{\left(y-x\right)}^2/4t}-\e^{-{\left(y+x\right)}^2/4t}}{\sqrt{4\pi t}}\right).
\end{eqnarray}
\endnumparts
Similar to the quenched ensemble, the solution for these perturbation fields are straightforward to write in terms of $g(x,y,t)$ in  \eref{green_function_arbitrary_boundary_rate} and $\widehat{g}(x,y,t)$ in \eref{eq:modified greens sol}. We present here only the results for the fast coupling limit, which yields simpler expression.

In the fast coupling limit, it is easier to see that the variance re-scaled by $\sqrt{T}$ differs from its expression \eref{eq:variance quench not explicit} in the quenched case by $\frac{2-\sqrt{2}}{\sqrt{\pi}}\; \sigma(\rho_b)$. This yields the explicit result \eref{eq:variance ann fast}. It is instructive to compare with a similar result \cite{krapivsky_fluctuations_2012} in the infinite geometry.

For the third cumulant in the fast coupling limit we find
\begin{eqnarray}
    \fl\frac{\left<Q_T^{\,3}\right>_{\mathcal{Q}}-\left<Q_T^{\,3}\right>_{\mathcal{A}}}{\sqrt{T}}=3\int_0^1\mathrm{d}t\int_0^{\infty}\mathrm{d}x\,\sigma'(\rho_{\mathrm{av}}(x,t))\frac{\e^{-x^2/2\left(1-t\right)}}{\pi\left(1-t\right)} \int_0^\infty\mathrm{d}y\,\psi_1(y,0)\partial_x\widehat{g}_{\rm fast}(x,y,t)\cr
    \fl\qquad-\int_0^\infty\mathrm{d}x\left(6\psi_2(x,0) \widehat{g}_{\rm fast}(x,0,1)-\frac{12}{\sigma(\rho_b)}q_1(x,0)q_2(x,0)+\frac{2\sigma'(\rho_b)}{\sigma^2(\rho_b)}q_1^3(x,0)\right)\label{eq:Q3 a-q formal}
\end{eqnarray}
where we have defined $\psi_{1(2)}(x,0)=-\int_0^x\mathrm{d}y\,q_{1(2)}(y,0)$ and
\begin{equation}
   \widehat{g}_{\rm fast}(x,y,t)=\frac{\e^{-{\left(y-x\right)}^2/4t}+\e^{-{\left(y+x\right)}^2/4t}}{\sqrt{4\pi t}}.
\end{equation}

The intricate expression \eref{eq:Q3 a-q formal}  reduces to the explicit result \eref{eq:Q3 ann fast}. We carried out this simplification in two steps. First, for $(\rho_a,\rho_b)\equiv(\frac{1}{2},0)$ where the quenched and the annealed results are identical, therefore the difference in \eref{eq:Q3 a-q formal} vanishes, $\left<Q_T^{\,3}\right>_{\mathcal{A}}$ is computed by comparing with similar analysis for the infinite line \cite{derrida_current_2009} and using the known result \cite{Gerschenfeld2009Bethe}. The expression \eref{eq:Q3 ann fast} for arbitrary densities is then obtained from the $(\frac{1}{2},0)$ result using a relation \cite{Gerschenfeld2009Bethe,Derrida2004Roche} for the annealed cgf between any two pair of values of $(\rho_a,\rho_b)$. The relation is a consequence of the symmetry discussed in the last remark in \sref{Sect: Macro}.

\section{Conclusion}\label{sec:concl}
In this article we examined a one dimensional semi-infinite SSEP connected to a particle reservoir at one end with a varied coupling strength. We showed (\sref{Sect: Macro}) how to apply the well-known MFT approach in this non-equilibrium context for studying current fluctuations in the large time limit. In the MFT formalism, the slow coupling with reservoir brings an additional term to the hydrodynamic Action. For the SSEP, we presented (\ref{Derive_Action_SSEP}) a derivation of this term which is simple to generalize for similar transport models. Using the MFT formalizm we found explicit expression for the cgf of current (\sref{sec:NI hydro}) in the low density limit. We confirmed the MFT results using an exact solution starting with the microscopic dynamics (\sref{Sect: Micro}).  Not-surprisingly the results show a long-term memory of the initial condition which was reported earlier in similar contexts. For arbitrary density we used a standard perturbation solution of the MFT to derive explicit results for the first few cumulants of the current (\sref{sec:sep hydro}). 

The perturbation solution of the MFT is a systematic way to obtain the cumulants order by order. Similar to the finite \cite{Derrida2004Roche} and the infinite \cite{derrida_current_2009} SSEP, the cgf of current for the annealed ensemble depends on $\rho_a$, $\rho_b$, and $\lambda$ through a single function $\omega=\rho_a\left(1-\rho_b\right)\left(\e^\lambda-1\right)+\rho_b\left(1-\rho_a\right)\left(\e^{-\lambda}-1\right)$. This indicates that the cgf has a series expansion in powers of $\omega$ which might be possible to predict using the perturbation solution.

The appealing idea about the hydrodynamic approach of the MFT is its generality. For the semi-infinite line with the slow coupling our presentation is easily extendable for other diffusive transport models. For example, the exclusion process with weak drift \cite{Enaud2004}, the KMP model of heat conduction \cite{bertini_large_2005}, the zero-range process \cite{Harris2005}, and the symmetric inclusion process \cite{Carinci2013} have a similar hydrodynamic description \cite{bertini_macroscopic_2015} and it would be of interest to obtain their explicit results for current fluctuations.

The aspect of integrability of the variational problem within the MFT has seen a remarkable progress in recent years \cite{Krajenbrink2021,Bettelheim2022,mallick2022exact}. On the infinite lattice, the variational problem for the SSEP has been solved \cite{mallick2022exact}. Corresponding solution for the semi-infinite geometry still poses a challenging open problem for the future. 

\ack
We acknowledge support of the Department of Atomic Energy, Government of India, under Project Identification No. RTI 4002. TS acknowledges insightful discussions with Bernard Derrida about the results in this work. Many crucial ideas, especially the importance of considering the slow coupling, the derivation of the hydrodynamic action, and the microscopic solution for the non-interacting case originated from those discussions. 
\appendix

\section{A derivation of the hydrodynamic-Action for the SSEP} \label{Derive_Action_SSEP}

We present a non-rigorous, but simple approach for deriving the fluctuating hydrodynamics description for SSEP. A similar analysis for stochastic processes was discussed in an earlier work of Lef\`evre and Biroli \cite{Lefevre2007}. We follow our notation in this article for the SSEP on a $\mathbb{Z}^+$  lattice with the sites indexed by  $i=1,\,2,\,3,\,\cdots,$ and the microscopic time denoted by $\tau$. Jump rates are given in \fref{fig: SSEP Slow}. A configuration of the system is specified by the set $\{n_i(\tau)\}$ where $n_i(\tau)$ denotes the occupancy of the $i$-th lattice site at a given instant of time $\tau$. Here, $n_i$ is equal to $0$ if the site is empty whereas it equals to $1$ if the site is filled with one particle. For the SSEP, there can be at most one particle in a site at a given time.

The total measurement time $T$ is broken into $M$ number of infinitesimal steps each of a duration $\mathrm{d}\tau$. This allows us to denote any instant of the microscopic time $\tau$ as $k\,\mathrm{d}\tau$ such that $k=0,\,1,\,2,\,\cdots,\,M$ with $M\,\mathrm{d}\tau=T$. 

The conservation law for the number of particles states that
\begin{equation}
    n_i(\tau+\mathrm{d}\tau)-n_i(\tau)=Y_{i-1}(\tau)-Y_i(\tau)\label{eq:conservation micro}
\end{equation}
where $Y_i(\tau)$ denotes the flux of particles in the infinitesimal time window between $\tau$ and $\tau+\mathrm{d}\tau$ across the bond connecting the $i$-th and the $\left(i+1\right)$-th sites. The $Y_0(\tau)$ denotes the influx of particles from the reservoir to the boundary site $i=1$.

In our notation, the net current of particles from the reservoir in time $T$ is $Q_T=\sum_{k=0}^{M-1}Y_0(k\,\mathrm{d}\tau)$ and using \eref{eq:conservation micro} we get
\begin{equation}
    Q_T=\sum_{i=1}^{\infty}\left(n_i(T)-n_i(0)\right).
\end{equation}
This is the microscopic analogue of \eref{time_int_current_defn}.

In an infinitesimal time step $\mathrm{d}\tau$, the $Y_i(\tau)$ can have three possible values: $0$ or $\pm 1$. In the bulk of the lattice (i.e., for $i\geq1$),
\begin{equation}
    \fl Y_{i}(\tau)=\cases{1&with prob. $n_i(\tau)\left(1-n_{i+1}(\tau)\right)\mathrm{d}\tau$\\
    -1&with prob. $n_{i+1}(\tau)\left(1-n_i(\tau)\right)\mathrm{d}\tau$\\
    0&with prob. $1 - \left[n_i(\tau)\left(1-n_{i+1}(\tau)\right)+  n_{i+1}(\tau)\left(1-n_i(\tau)\right)\right]\mathrm{d}\tau$.}
    \label{eq:Q prob bulk}
\end{equation}
Across the system-reservoir link
\begin{equation}
    \fl Y_{0}(\tau)=\cases{1&with prob. $\gamma\,\rho_a\left(1-n_1(\tau)\right)\mathrm{d}\tau$\\
    -1&with prob. $\gamma\,n_1(\tau)\left(1-\rho_a\right)\mathrm{d}\tau$\\
    0&with prob. $1-\gamma\left[\rho_a\left(1-n_1(\tau)\right)+ n_1(\tau)\left(1-\rho_a\right)\right]\mathrm{d}\tau$.}\label{eq:Q prob boundary}
\end{equation}

The generating function $\left<e^{\lambda Q_T}\right>_{\mathrm{hist}}$ where the average is over all history starting with a fixed initial configuration $\{n_i(0)\}$ is
\begin{eqnarray}
    \fl\qquad\quad\left<e^{\lambda Q_T}\right>_{\mathrm{hist}}=\int\mathcal{D}\left[n,Y\right]&\left< \prod_{i=1}^{\infty}e^{\lambda \left[n_i(T)-n_i(0)\right]}\right.\nonumber\\
    \fl\qquad\quad&\quad\left.\times\prod_{k=0}^{M-1}\delta_{n_i(k\,\mathrm{d}\tau+\mathrm{d}\tau)-n_i(k\,\mathrm{d}\tau)\,,\,Y_{i-1}(k\,\mathrm{d}\tau)-Y_i(k\,\mathrm{d}\tau)}\right>_{\left[Y\right]}
\end{eqnarray}
where the path integral measures are defined as
\begin{equation}
    \int\mathcal{D}\left[n\right]=\prod_{k=1}^{M}\prod_{i=1}^{\infty}\sum_{n_i(k\,\mathrm{d}\tau)=0}^1\;\;\mathrm{and}\;\;\int\mathcal{D}\left[Y\right]=\prod_{k=0}^{M-1}\prod_{i=0}^{\infty}\sum_{Y_i(k\,\mathrm{d}\tau)=-1}^1
\end{equation}
The $\langle \rangle_{\left[Y\right]}$ denotes average over all evolution of $\{Y_i(\tau)\}$ with appropriate probability weight.

Using an integral representation $\delta_{a,b}=\frac{1}{2\pi\mathrm{i}}\int_{-\mathrm{i}\pi}^{\mathrm{i}\pi}\mathrm{d}z\,\e^{-z\left(a-b\right)}$ for integers $a$ and $b$, we rewrite the expression
\begin{eqnarray}
    \fl\quad\left<\e^{\lambda Q_T}\right>_{\mathrm{hist}}=\int\mathcal{D}\left[n,\widehat{n}\right]&\e^{\lambda\sum_{i=1}^{\infty}\big(n_i(T)-n_i(0)\big)}\e^{-\sum_{k=0}^{M-1}\sum_{i=1}^{\infty}\widehat{n}_i(k\,\mathrm{d}\tau)\big(n_i(k\,\mathrm{d}\tau+\mathrm{d}\tau)-n_i(k\,\mathrm{d}\tau)\big)}\nonumber\\
    \fl\quad&\times\int\mathcal{D}\left[Y\right]\left<\e^{\sum_{k=0}^{M-1}\sum_{i=1}^{\infty}\widehat{n}_i(k\,\mathrm{d}\tau)\big(Y_{i-1}(k\,\mathrm{d}\tau)-Y_i(k\,\mathrm{d}\tau)\big)}\right>_{[Y]}\label{prob_2_parts}
\end{eqnarray}
with the path integral measure on $\widehat{n}$ defined as
\begin{equation}
    \int\mathcal{D}\left[\widehat{n}\right]=\prod_{k=0}^{M-1}\prod_{i=1}^\infty \frac{1}{2\pi\mathrm{i}} \int_{-\mathrm{i}\pi}^{\mathrm{i}\pi}\mathrm{d}\widehat{n}_i(k\,\mathrm{d}\tau)
\end{equation}

To compute the average over $Y$ we rearrange the terms in the exponential
\begin{eqnarray}
    \fl\sum_{i=1}^{\infty}\widehat{n}_i(k\,\mathrm{d}\tau)\big(Y_{i-1}(k\,\mathrm{d}\tau)-Y_i(k\,\mathrm{d}\tau)\big)\nonumber\\
    =\widehat{n}_1(k\,\mathrm{d}\tau)Y_0(k\,\mathrm{d}\tau)+ \sum_{i=1}^{\infty}\big(\widehat{n}_{i+1}(k\,\mathrm{d}\tau)-\widehat{n}_i(k\,\mathrm{d}\tau)\big)Y_i(k\,\mathrm{d}\tau)
\end{eqnarray}
and write
\begin{eqnarray}
    \fl \left<\e^{\sum_{k=0}^{M-1}\sum_{i=1}^{\infty}\left[\widehat{n}_i(k\,\mathrm{d}\tau)\big(Y_{i-1}(k\,\mathrm{d}\tau)-Y_i(k\,\mathrm{d}\tau)\big)\right]}\right>_{[Y]} \nonumber\\
   \fl  =\prod_{k=0}^{M-1}\bigg<\exp\left(\widehat{n}_1(k\,\mathrm{d}\tau)Y_0(k\,\mathrm{d}\tau)\right)\bigg>_{Y_0}\prod_{i=1}^{\infty}\bigg<\exp\left[\left(\widehat{n}_{i+1}(k\,\mathrm{d}\tau)-\widehat{n}_i(k\,\mathrm{d}\tau)\right)Y_i(k\,\mathrm{d}\tau)\right]\bigg>_{\left[Y_i\right]}
\end{eqnarray}
For the bulk sites ($i>0$) using the probability weight in \eref{eq:Q prob bulk} we get
\begin{eqnarray}
    \fl\qquad&\bigg<\exp\left[\left(\widehat{n}_{i+1}(k\,\mathrm{d}\tau)-\widehat{n}_i(k\,\mathrm{d}\tau)\right)Y_i(k\,\mathrm{d}\tau)\right]\bigg>_{\left[Y_i\right]}\nonumber\\ \fl\qquad=&\exp\bigg\{\mathrm{d}\tau\Big[\left(\e^{\widehat{n}_{i+1}(k\,\mathrm{d}\tau)-\widehat{n}_i(k\,\mathrm{d}\tau)}-1\right)n_i(k\,\mathrm{d}\tau)\left(1-n_{i+1}(k\,\mathrm{d}\tau)\right)\cr
    \fl\qquad&\quad\qquad\quad+\left(\e^{\widehat{n}_i(k\,\mathrm{d}\tau)-\widehat{n}_{i+1}(k\,\mathrm{d}\tau)}-1\right)n_{i+1}(k\,\mathrm{d}\tau)\left(1-n_i(k\,\mathrm{d}\tau)\right)\Big]\bigg\}\label{eq:av bulk}
\end{eqnarray}
and for the boundary using \eref{eq:Q prob boundary} we get
\begin{eqnarray}
    \fl\;\bigg<\exp\left(\widehat{n}_1(k\,\mathrm{d}\tau)Y_0(k\,\mathrm{d}\tau)\right)\bigg>_{Y_0} =\exp\bigg\{\gamma\,\mathrm{d}\tau\Big[&\left(\e^{\widehat{n}_1(k\,\mathrm{d}\tau)}-1\right)\rho_a\left(1-n_1(k\,\mathrm{d}\tau)\right)\nonumber\\
    \fl\;&+\left(\e^{-\widehat{n}_1(k\,\mathrm{d}\tau)}-1\right)n_1(k\,\mathrm{d}\tau)\left(1-\rho_a\right)\Big]\bigg\}.\label{eq:av bndry}
\end{eqnarray}

Using (\ref{eq:av bulk},~\ref{eq:av bndry}) in \eref{prob_2_parts} and taking the continuous time limit $\mathrm{d}\tau\to0$, we get
\begin{eqnarray}
    \fl\left<e^{\lambda Q_T}\right>_{\mathrm{hist}}=\int\mathcal{D}\left[n,\widehat{n}\right]&\exp\left[\lambda \sum_{i=1}^{\infty}\left(n_i(T)-n_i(0)\right)\right]\nonumber\\
    \fl&\exp\Bigg\{-\int_0^T\mathrm{d}\tau\Bigg[\left(\sum_{i=1}^\infty\widehat{n}_i(\tau)\frac{\mathrm{d}n_i(\tau)}{\mathrm{d}\tau}\right)-\mathcal{H}\left[n(\tau),\widehat{n}(\tau)\right]\Bigg]\Bigg\}\label{eq:micro action}
\end{eqnarray}
where the effective Hamiltonian $\mathcal{H}=\mathcal{H}_{\mathrm{bulk}}+\mathcal{H}_{\mathrm{bdry}}$ with the bulk contribution is given by
\begin{eqnarray}
    \mathcal{H}_{\mathrm{bulk}}=\sum_{i=1}^\infty&\bigg[n_i(\tau)\left(1-n_{i+1}(\tau)\right)\left(\e^{\widehat{n}_{i+1}(\tau)-\widehat{n}_i(\tau)}-1\right)\nonumber\\
    &+n_{i+1}(\tau)\left(1-n_i(\tau)\right)\left(\e^{\widehat{n}_i(\tau)-\widehat{n}_{i+1}(\tau)}-1\right)\bigg]
\end{eqnarray}
and the effective boundary Hamiltonian is
\begin{equation}
    \fl\qquad\qquad\mathcal{H}_{\mathrm{bdry}}=\gamma\left[\rho_a\left(1-n_1(\tau)\right)\left(\e^{\widehat{n}_1(\tau)}-1\right)+n_1(\tau)\left(1-\rho_a\right)\left(\e^{-\widehat{n}_1(\tau)}-1\right)\right].
\end{equation}

In the macroscopic scale $(x,t)\equiv\left(\frac{i}{\sqrt{T}},\,\frac{\tau}{T}\right)$ for large $T$, assuming convergence to slowly varying hydrodynamic fields $n_i(\tau)\to\rho(x,t)$ and $\widehat{n}_i(\tau)\to\widehat{\rho}(x,t)$, and following gradient expansion we arrive at the leading asymptotic of the generating function for large $T$, given in \eref{action_hamilton_defn} for the case $\mathcal{F}=0$.

\begin{remark}
In taking the continuous time limit in \eref{eq:micro action} we have written
\begin{eqnarray}
    \sum_{k=0}^{M-1}\widehat{n}_i(k\,\mathrm{d}\tau)\left[n_i(k\,\mathrm{d}\tau+\mathrm{d}\tau)-n_i(k\,\mathrm{d}\tau)\right]
    \to \int_0^T\mathrm{d}\tau\; \widehat{n}_i(\tau)\frac{\mathrm{d}n_i(\tau)}{\mathrm{d}\tau}.\label{prob_1st_part}
\end{eqnarray}
Considering that $n_i$ is either $0$ or $1$, the $\frac{\mathrm{d}n_i(\tau)}{\mathrm{d}\tau}$ is ill defined. An interpretation of \eref{eq:micro action} may be drawn by defining the probability measure with respect to another simpler process. For example, a particle deposition-evaporation process on the semi-infinite lattice with unit rate and simple exclusion has similar path probability with an effective Hamiltonian
\begin{equation}
 \mathcal{H}_{\mathrm{DE}}\left[n_i(\tau),\widehat{n}_i(\tau)\right]=\left(1-n_{i}(\tau)\right)\left(\e^{\widehat{n}_i(\tau)}-1\right)+n_{i}(\tau)\left(\e^{-\widehat{n}_{i}(\tau)}-1\right).
\end{equation}
Then the Radon-Nikodym derivative \cite{Touchette2018} of the path-probability for SSEP with respect to the deposition-evaporation process is well-defined.
\end{remark}

\subsection{Non-Interacting particles} \label{Derive_Action_NIRW}

We proceed in the same manner as we did for the SSEP in the previous section. The main difference comes in the probabilities with which the particle flux takes the possible values.

The particle flux for the bulk lattice sites i.e., $i\geq1$ at any time $\tau$ can have three possible values with certain probabilities given by
\begin{equation}
    \fl\qquad Y_{i}(\tau)=\cases{1&with probability $n_i(\tau)\,\mathrm{d}\tau$\\
    -1&with probability $n_{i+1}(\tau)\,\mathrm{d}\tau$\\
    0&with probability $1-\left(n_i(\tau)+n_{i+1}(\tau)\right)\mathrm{d}\tau$.}
\end{equation}
On the other hand, across the system-reservoir boundary i.e., for $i=0$ we have
\begin{equation}
    \fl\qquad Y_{0}(\tau)=\cases{1&with probability $\alpha\,\mathrm{d}\tau=\gamma\,\rho_a\,\mathrm{d}\tau$\\
    -1&with probability $\gamma\,n_1(\tau)\,\mathrm{d}\tau$\\
    0&with probability $1-\gamma\left(\rho_a+n_1(\tau)\right)\mathrm{d}\tau$.}
\end{equation}

Following a similar analysis as presented for the SSEP we arive at the cgf \eref{eq:micro action} with 
\begin{eqnarray}    \mathcal{H}_{\mathrm{bulk}}\left[n,\widehat{n}\right]=\left(\e^{\widehat{n}_{i+1}(\tau)-\widehat{n}_i(\tau)}-1\right)n_i(\tau)+\left(\e^{\widehat{n}_i(\tau)-\widehat{n}_{i+1}(\tau)}-1\right)n_{i+1}(\tau),\\
    \mathcal{H}_{\mathrm{bdry}}\left[n,\widehat{n}\right]=\gamma\left[\left(\e^{\widehat{n}_1(\tau)}-1\right)\rho_a+\left(\e^{-\widehat{n}_1(\tau)}-1\right)n_1\right].
\end{eqnarray}

In the hydrodynamic limit, this leads to the cgf in \eref{time_int_current_defn}-\eref{eq:Htotal} with $D(q)=1$ and $\sigma(q)=2q$ and the boundary hamiltonian \eref{boundary_action_nirw}.

\subsection{ Fluctuating hydrodynamics} \label{Derive_fld_SEP}
The Action \eref{action_hamilton_defn} for $\lambda=0$ and $\mathcal{F}=0$ is the Martin-Siggia-Rose-Janssen-De Dominicis action \cite{msrd_1,msrd_2,msrd_3,msrd_4} for the fluctuating hydrodynamics equation (\ref{eq:flhd full},\,\ref{eq:flhd boundary}) where $\widehat{\rho}$ is the response field. One simple way to see this is by writing the cgf of $Q_T$ in \eref{time_int_current_defn} for (\ref{eq:flhd full},\,\ref{eq:flhd boundary}).
\begin{eqnarray}
    \fl\qquad\langle e^{\lambda Q_T} \rangle=\int \mathcal{D}\left[\rho\right]&\e^{\lambda\sqrt{T}\int_0^\infty\mathrm{d}x\left(\rho(x,1)-\rho(x,0)\right)}\bigg\langle\delta\left(\partial_x\rho(0,t)+\eta(0,t)+\xi(t)\right)\nonumber\\
    \fl\qquad&\qquad\qquad\prod_{x>0}\delta\left(\partial_t\rho(x,t)-\partial_{xx}\rho(x,t)-\partial_x\eta(x,t)\right)\bigg\rangle_{\xi,\,\eta} 
\end{eqnarray}
where the angular brackets denote average over realizations of the noises $\xi(t)$ and $\eta(x,t)$. 
Using an integral representation of the delta function $\delta(x)=\left(2\pi\mathrm{i}\right)^{-1}\int_{-\mathrm{i}\pi}^{\mathrm{i}\pi}\mathrm{d}z\,\e^{xz}$ to introduce the response field $\widehat{\rho}(x,t)$ and subsequently following integration by parts to compute the averages over noise realizations lead to the path integral 
\eref{mgf_path_integral} with $\mathcal{F}=0$. 

A more careful analysis can be done by discretizing the (\ref{eq:flhd full},\,\ref{eq:flhd boundary}) following It\^{o} convention, which confirms the expression \eref{action_hamilton_defn}.

\section{The Green's function}\label{green_function_details}
The Green's function $g(x,y,t)$ in \eref{green_function_arbitrary_boundary_rate} is the solution of the diffusion equation
\begin{equation}
    \frac{\partial g(x,y,t)}{\partial t}=\frac{\partial^2g(x,y,t)}{\partial x^2}\label{diffusion_eqn_green_func}
\end{equation}
on the semi-infinite line $x>0$ with the spatial boundary condition
\begin{eqnarray}
    g(x,y,t)-\frac{1}{\Gamma}\frac{\partial g(x,y,t)}{\partial x}=0\quad\mathrm{at}\;x=0\quad\mathrm{and}\label{left_spatial_boundary_green_func}\\
    g(x,y,t)=0\quad\mathrm{at}\;x\to\infty\label{right_spatial_boundary_green_func}
\end{eqnarray}
and the initial condition
\begin{equation}
    g(x,y,0)=\delta(x-y).\label{initial_cond_green_func}
\end{equation}
We have derived the solution for the Green's function in \eref{green_function_arbitrary_boundary_rate} using the method of spectral decomposition of the diffusion operator. Alternatively, one can make a change of variables as $\mathcal{G}(x,y,t)=g(x,y,t)-\Gamma^{-1}\partial_xg(x,y,t)$ such that $\mathcal{G}(x,y,t)$ now satisfies a heat equation on the semi-infinite line with a Dirichlet boundary condition. To get back the desired solution of $g(x,y,t)$, one then needs to solve a self-consistent linear ode with appropriate boundary conditions.

We have not been able to complete the integration in \eref{green_function_arbitrary_boundary_rate} over the $z$-variable for arbitrary $\Gamma$. In the two limiting cases, the Green's function has an explicit expression.
\begin{equation}
     g(x,y,t)=\frac{\e^{-{\left(y-x\right)}^2/4t}-\e^{-{\left(y+x\right)}^2/4t}}{\sqrt{4\pi t}}\qquad \textrm{for $\Gamma\to \infty$}\label{green_func_strong_fast_limit}
\end{equation}
which satisfies diffusion equation with absorbing boundary at $x=0$, and 
\begin{equation}
     g(x,y,t)=\frac{\e^{-{\left(y-x\right)}^2/4t}+\e^{-{\left(y+x\right)}^2/4t}}{\sqrt{4\pi t}}\qquad \textrm{for $\Gamma\to 0$}\label{green_func_weak_slow_limit}
\end{equation}
which satisfies diffusion equation with reflecting boundary at $x=0$.

The solution of the perturbation fields $q_1$ and $q_2$ in \eref{eq:perturbation of fields} are expressed in terms of a second Green's function $\widehat{g}(x,y,t)$, which is the solution of the diffusion equation \eref{diffusion_eqn_green_func} with the difference in the boundary condition
\begin{equation}
    \frac{\partial\widehat{g}(x,y,t)}{\partial x}-\frac{1}{\Gamma}\frac{\partial^2\widehat{g}(x,y,t)}{\partial x^2}=0\quad\mathrm{at}\;x=0.\label{left_spatial_boundary_modified_green_func}
\end{equation}
The corresponding solution is
\begin{equation}
    \fl\widehat{g}(x,y,t)=2\int_0^\infty\mathrm{d}z\,\Bigg[\frac{\e^{-\pi^2tz^2}}{\Gamma^2+\pi^2z^2}\left(\Gamma\cos\pi xz-\pi z\sin\pi xz\right)\left(\Gamma\cos\pi yz-\pi z\sin\pi yz\right)\Bigg].\label{eq:modified greens sol}
\end{equation}

\section{Derivation of the identity \eref{space_int_final_time_int_origin_green}}
Using \eref{diffusion_eqn_green_func}, we write
\begin{equation}
    \frac{\partial}{\partial t}\left(1-\int_0^\infty\mathrm{d}y\,g(x,y,1-t)\right)=\frac{\partial^2}{\partial x^2}\int_0^\infty\mathrm{d}y\,g(x,y,1-t).
\end{equation}
Integrating both sides over $t\in[0,1]$, we get
\begin{equation}
    -\left(1-\int_0^\infty\mathrm{d}y\,g(x,y,1)\right)=\frac{\partial^2}{\partial x^2}\int_0^1\mathrm{d}t\int_0^\infty\mathrm{d}y\,g(x,y,1-t)
\end{equation}
where we have used \eref{initial_cond_green_func} so that the LHS vanishes at $t=1$. Integrating over $x\in[0,\infty)$, we obtain
\begin{eqnarray}
    \fl-\int_0^\infty\mathrm{d}x\left(1-\int_0^\infty\mathrm{d}y\,g(x,y,1)\right)=&\left.\left(\frac{\partial}{\partial x}\int_0^1\mathrm{d}t\int_0^\infty\mathrm{d}y\,g(x,y,t)\right)\right|_{x=\infty}\nonumber\\
    \fl&-\left.\left(\frac{\partial}{\partial x}\int_0^1\mathrm{d}t\int_0^\infty\mathrm{d}y\,g(x,y,t)\right)\right|_{x=0}.
\end{eqnarray}
Using the boundary condition (\ref{left_spatial_boundary_green_func},\,\ref{right_spatial_boundary_green_func}) the above reduces to \eref{space_int_final_time_int_origin_green}.

\section{The average density profile}\label{app:derivation ave rho}

Recently, there have been much interest \cite{Baldasso2017,Franco2019,Franco2016,Goncalves2018,Tsunoda2019,franceschini_hydrodynamical_2022,franceschini_symmetric_2021,Derrida2021} in the effect of slow coupling to a reservoir on the hydrodynamics for the average density of the SSEP. Here we present a derivation of (\ref{average_profile_diffusion_equation_transport_coefficient},~\ref{avg_density_boundary_condition}) for the SSEP on a semi-infinite lattice $i\in\mathbb{Z}^+$ illustrated in \fref{fig: SSEP Slow}.

The average occupancy $\left<n_i(\tau)\right>$ of the $i$-th lattice site at time $\tau$ evolves according to the rate equation
\begin{equation}
    \frac{\mathrm{d}\left<n_i(\tau)\right>}{\mathrm{d}\tau}=\cases{-\left(1+\gamma\right)\left<n_1(\tau)\right>+\left<n_2(\tau)\right>+\gamma\rho_a&for $i=1$,\\
    \left<n_{i-1}(\tau)\right>-2\left<n_i(\tau)\right>+\left<n_{i+1}(\tau)\right>&for $i\geq2$.\\}
\end{equation}

At $\tau\to\infty$, the average occupation asymptotically reaches an equilibrium stationary state where for each site, it becomes equal to the reservoir density $\rho_a$. For convenience of algebra, we write
\begin{equation}
    \left<n_i(\tau)\right>=\rho_a+\left<\delta n_i(\tau)\right>
\end{equation}
such that the time dependent part follows a linear homogeneous equation
\begin{equation}
     \frac{\mathrm{d}\left<\delta n_i(\tau)\right>}{\mathrm{d}\tau}=\cases{-\left(1+\gamma\right)\left<\delta n_1(\tau)\right>+\left<\delta n_2(\tau)\right>&for $i=1$,\\
    \left<\delta n_{i-1}(\tau)\right>-2\left<\delta n_i(\tau)\right>+\left<\delta n_{i+1}(\tau)\right>&for $i\geq2$.\\}\label{eq:for delta n}
\end{equation}

A formal solution of \eref{eq:for delta n} is
\begin{equation}
    \left<\delta n_i(\tau)\right>=2\int_{-\pi}^{\pi}\frac{\mathrm{d}\theta}{2\pi}\,\e^{-\tau\,2\left(1-\cos\theta\right)}\frac{\Psi_i(\theta)\sum_{j=1}^\infty\Psi_j(\theta)\left<\delta n_j(0)\right>}{1+\left(\gamma-1\right)^2+2\left(\gamma-1\right)\cos\theta}
\end{equation}
where
\begin{equation}
    \Psi_i(\theta)=\sin\left(\theta i\right)+\left(\gamma-1\right)\sin\left[\theta\left(i-1\right)\right].\label{normalised_eigenvectors}
\end{equation}

To make the summation over $j$ finite we choose an initial condition $\left<\delta n_j(0)\right>=\rho_b-\rho_a$ for $j\le L$ and $0$ for the rest. This gives
\begin{equation}
    \sum_{j=1}^L\Psi_j(\theta)\left<\delta n_j(0)\right>=\left(\rho_b-\rho_a\right)\left[\sin L\theta+\gamma\frac{\sin\frac{L\theta}{2}\,\sin\frac{\left(L-1\right)\theta}{2}}{\sin\frac{\theta}{2}}\right].
\end{equation}

For $L^2\gg\tau$ the evolution near the reservoir approximates the evolution with uniform initial density throughout the infinite lattice. In the hydrodynamic limit for large $T$ with $\gamma=\frac{\Gamma}{\sqrt{T}}$ and $L=\alpha \sqrt{T}$, the solution $\left<\delta n_i(\tau)\right>\simeq \left<\delta\rho\left(\frac{i}{\sqrt{T}},\frac{\tau}{T}\right)\right>$ with
\begin{eqnarray}
    \fl\qquad\qquad\left<\delta\rho(x,t)\right>=2\Gamma\left(\rho_b-\rho_a\right)\int_0^\infty&\frac{\mathrm{d}z}{\pi z}\,\frac{\e^{-t\pi^2z^2}}{\left(\Gamma^2+\pi^2z^2\right)}\left(\Gamma\sin\pi xz+\pi z\cos\pi xz\right)\nonumber\\
    &\;\left(1-\cos\pi z\alpha+\frac{\pi z}{\Gamma}\sin\pi z\alpha\right)\label{eq:delta rho}
\end{eqnarray}
where we made a change of variable $\theta=\frac{\pi z}{\sqrt{T}}$. For large $\alpha$ the trigonometric functions involving $\alpha$ oscillates fast and their contribution is sub-leading. This leads to the expression of the average density $\rho_{\mathrm{av}}(x,t)=\rho_a+\left<\delta\rho(x,t)\right>$ in \eref{eq:rhoave sol}.

It is now straightforward to verify that the solution \eref{eq:rhoave sol} satisfies the diffusion equation \eref{average_profile_diffusion_equation_transport_coefficient} with the Robin boundary condition \eref{avg_density_boundary_condition}.

\section*{References}
\bibliographystyle{iopart-num}
\bibliography{references}

\providecommand{\newblock}{}
\begin{thebibliography}{10}
\expandafter\ifx\csname url\endcsname\relax
  \def\url#1{{\tt #1}}\fi
\expandafter\ifx\csname urlprefix\endcsname\relax\def\urlprefix{URL }\fi
\providecommand{\eprint}[2][]{\url{#2}}

\bibitem{bertini_macroscopic_2015}
Bertini L, De~Sole A, Gabrielli D, Jona-Lasinio G and Landim C 2015 {\em {Rev.
  Mod. Phys.}\/} {\bf 87} 593

\bibitem{derrida_microscopic_2011}
Derrida B 2011 {\em {J. Stat. Mech.}\/}  P01030

\bibitem{bertini_fluctuations_2001}
Bertini L, De~Sole A, Gabrielli D, Jona-Lasinio G and Landim C 2001 {\em {Phys.
  Rev. Lett.}\/} {\bf 87} 040601

\bibitem{bertini_macroscopic_2002}
Bertini L, De~Sole A, Gabrielli D, Jona-Lasinio G and Landim C 2002 {\em {J.
  Stat. Phys.}\/} {\bf 107} 635

\bibitem{derrida_large_2002}
Derrida B, Lebowitz J~L and Speer E~R 2002 {\em {J. Stat. Phys.}\/} {\bf 107}
  599

\bibitem{bodineau_current_2004}
Bodineau T and Derrida B 2004 {\em {Phys. Rev. Lett.}\/} {\bf 92} 180601

\bibitem{Derrida2004Roche}
Derrida B, Dou{\c{c}}ot B and Roche P~E 2004 {\em {J. Stat. Phys.}\/} {\bf 115}
  717

\bibitem{tailleur_mapping_2007}
Tailleur J, Kurchan J and Lecomte V 2007 {\em {Phys. Rev. Lett.}\/} {\bf 99}
  150602

\bibitem{bertini_current_2005}
Bertini L, De~Sole A, Gabrielli D, Jona-Lasinio G and Landim C 2005 {\em {Phys.
  Rev. Lett.}\/} {\bf 94} 030601

\bibitem{Lecomte_Current_2010}
Lecomte V, Imparato A and van Wijland F 2010 {\em Prog. Theor. Phys. Suppl.\/}
  {\bf 184} 276

\bibitem{Gerschenfeld2009Bethe}
Derrida B and Gerschenfeld A 2009 {\em {J. Stat. Phys.}\/} {\bf 136} 1

\bibitem{mallick2022exact}
Mallick K, Moriya H and Sasamoto T 2022 {\em {Phys. Rev. Lett.}\/} {\bf 129}
  040601

\bibitem{derrida_current_2009}
Derrida B and Gerschenfeld A 2009 {\em {J. Stat. Phys.}\/} {\bf 137} 978

\bibitem{bertini_large_2005}
Bertini L, Gabrielli D and Lebowitz J~L 2005 {\em {J. Stat. Phys.}\/} {\bf 121}
  843

\bibitem{Krapivsky2014}
Krapivsky P~L, Mallick K and Sadhu T 2014 {\em {Phys. Rev. Lett.}\/} {\bf 113}
  078101

\bibitem{Krapivsky2015}
Krapivsky P~L, Mallick K and Sadhu T 2015 {\em {J. Stat. Phys.}\/} {\bf 160}
  885

\bibitem{Krajenbrink2021}
Krajenbrink A and Le~Doussal P 2021 {\em {Phys. Rev. Lett.}\/} {\bf 127} 064101

\bibitem{Meerson2016}
Meerson B, Katzav E and Vilenkin A 2016 {\em {Phys. Rev. Lett.}\/} {\bf 116}
  070601

\bibitem{Liggett_1975}
Liggett T~M 1975 {\em {Trans. Am. Math. Soc.}\/} {\bf 213} 237

\bibitem{Grosskinsky_thesis}
Grosskinsky S 2004 {\em Phase transitions in nonequilibrium stochastic particle
  systems with local conservation laws\/} Ph.D. thesis Technical University of
  Munich

\bibitem{Williams2000}
Williams L and Sasamoto T 2012 Combinatorics of the asymmetric exclusion
  process on a semi-infinite lattice (\textit{Preprint}
  \eprint{arXiv:1204.1114})

\bibitem{Duhart_2018}
Duhart H~G, M{\"{o}}rters P and Zimmer J 2018 {\em {Potential Anal.}\/} {\bf
  48} 301

\bibitem{Tracy_2013}
Tracy C~A and Widom H 2013 {\em {J. Math. Phys.}\/} {\bf 54} 103301

\bibitem{Krapivsky2012}
Krapivsky P~L 2012 {\em {Phys. Rev. E}\/} {\bf 86} 1

\bibitem{Derrida2021}
Derrida B, Hirschberg O and Sadhu T 2021 {\em {J. Stat. Phys.}\/} {\bf 182} 15

\bibitem{Baldasso2017}
Baldasso R, Menezes O, Neumann A and Souza R~R 2017 {\em {J. Stat. Phys.}\/}
  {\bf 167} 1112

\bibitem{Franco2019}
Franco T, Gon{\c{c}}alves P and Neumann A 2019 {\em {Stoch. Process. Their
  Appl.}\/} {\bf 129} 1413

\bibitem{Franco2016}
Franco T, Gon{\c{c}}alves P and Neumann A 2017 Equilibrium fluctuations for the
  slow boundary exclusion process {\em Springer Proceedings in Mathematics and
  Statistics\/} vol 209 (Springer) p 177

\bibitem{Goncalves2018}
Gon{\c{c}}alves P, Jara M, Menezes O and Neumann A 2020 {\em {Stoch. Process.
  Their Appl.}\/} {\bf 130} 4326

\bibitem{Tsunoda2019}
Tsunoda K 2019 Hydrostatic limit for exclusion process with slow boundary
  revisited (\textit{Preprint} \eprint{arXiv:1902.06210v2})

\bibitem{franceschini_hydrodynamical_2022}
Franceschini C, Gonçalves P and Salvador B 2023 {\em Math. Phys. Anal.
  Geom.\/} {\bf 26} 11

\bibitem{franceschini_symmetric_2021}
Franceschini C, Gonçalves P and Sau F 2022 {\em {Bernoulli}\/} {\bf 28} 1340

\bibitem{Bodineau2009}
Bodineau T and Lagouge M 2010 {\em {J. Stat. Phys.}\/} {\bf 139} 201

\bibitem{DeMasi2011}
{De Masi} A, Presutti E, Tsagkarogiannis D and Vares M~E 2011 {\em {J. Stat.
  Phys.}\/} {\bf 144} 1151

\bibitem{DeMasi2012}
{De Masi} A, Presutti E, Tsagkarogiannis D and Vares M~E 2012 {\em {J. Stat.
  Phys.}\/} {\bf 147} 519

\bibitem{Franco2013}
Franco T, Gon{\c{c}}alves P and Neumann A 2013 {\em {Ann. Inst. H. Poincaré
  Probab. Statist.}\/} {\bf 49} 402

\bibitem{Redig2011}
Redig F and Vafayi K 2011 {\em {J. Math. Phys.}\/} {\bf 52} 093303

\bibitem{Landim2018}
Landim C and Tsunoda K 2018 {\em {Ann. Inst. H. Poincaré Probab. Statist.}\/}
  {\bf 54} 51

\bibitem{Erignoux2019}
Erignoux C, Gon{\c{c}}alves P and Nahum G 2020 {\em {J. Stat. Phys.}\/} {\bf
  181} 1433

\bibitem{tailleur_mapping_2008}
Tailleur J, Kurchan J and Lecomte V 2008 {\em {J. Phys. A: Math. Theor.}\/}
  {\bf 41} 505001

\bibitem{msrd_1}
Martin P~C, Siggia E~D and Rose H~A 1973 {\em {Phys. Rev. A}\/} {\bf 8} 423

\bibitem{msrd_2}
Janssen H~K 1976 {\em Z. Physik. B\/} {\bf 23} 377

\bibitem{msrd_3}
De~Dominicis C 1976 {\em {J. Phys. Colloques}\/} {\bf 37} C1--247

\bibitem{msrd_4}
De~Dominicis C and Peliti L 1978 {\em {Phys. Rev. B}\/} {\bf 18} 353

\bibitem{derrida_non-equilibrium_2007}
Derrida B 2007 {\em {J. Stat. Mech.}\/}  P07023

\bibitem{Derrida_2019}
Derrida B and Sadhu T 2019 {\em J Stat Phys\/} {\bf 177} 151

\bibitem{Rana_2023}
Rana J and Sadhu T 2023 {\em Phys. Rev. E\/} {\bf 107} L012101

\bibitem{sadhu_large_2015}
Sadhu T and Derrida B 2015 {\em {J. Stat. Mech.}\/}  P09008

\bibitem{krapivsky_fluctuations_2012}
Krapivsky P~L and Meerson B 2012 {\em {Phys. Rev. E}\/} {\bf 86} 031106

\bibitem{KrapivskyMelting2015}
Krapivsky P~L, Mallick K and Sadhu T 2015 {\em {J. Phys. A: Math. Theor.}\/}
  {\bf 48} 015005

\bibitem{Enaud2004}
Enaud C and Derrida B 2004 {\em J. Stat. Phys.\/} {\bf 114} 537

\bibitem{Harris2005}
Harris R~J, Rákos A and Schütz G~M 2005 {\em J. Stat. Mech.\/}  P08003

\bibitem{Carinci2013}
Carinci G, Giardinà C, Giberti C and Redig F 2013 {\em J. Stat. Phys.\/} {\bf
  152} 657

\bibitem{Bettelheim2022}
Bettelheim E, Smith N~R and Meerson B 2022 {\em Phys. Rev. Lett.\/} {\bf 128}
  130602

\bibitem{Lefevre2007}
Lef{\`{e}}vre A and Biroli G 2007 {\em {J. Stat. Mech.}\/}  P07024

\bibitem{Touchette2018}
Touchette H 2018 {\em J. Stat. Phys.\/} {\bf 170} 962

\end{thebibliography}

\end{document}